\documentclass[aps,prb,showpacs,reprint]{revtex4-1}

\usepackage{amsfonts}
\usepackage{amsmath}
\usepackage{amssymb}
\usepackage{graphicx}
\usepackage{subfigure}
\usepackage[utf8x]{inputenc}
\usepackage{color}
\usepackage{colortbl}
\usepackage[T1]{fontenc} 

\begin{document}

\title{Spaser and optical amplification conditions in gold-coated active nanoparticles.}

\author{Nicol\'{a}s Passarelli${}^{1}$, Ra\'{u}l A. Bustos-Mar\'{u}n${}^{1,2}$, Eduardo A. Coronado${}^{1}$}

\affiliation{${}^{1}$INFIQC-CONICET, Departamento de Fisicoqu\'{\i}mica, Facultad Ciencias Qu\'{\i}micas, UNC, Ciudad Universitaria, 5000 C\'{o}rdoba, Argentina, ${}^{2}$ IFEG-CONICET, Facultad de Matem\'{a}tica Astronom\'{i}a y F\'{i}sica, UNC, Ciudad Universitaria, 5000 C\'{o}rdoba, Argentina}

\email{coronado@fcq.unc.edu.ar}

\begin{abstract}
Due to their many potential applications, there is an increasing interest in studying hybrid systems composed of optically active media and plasmonic metamaterials. In this work we focus on a particular system which consists of an optically active silica core covered by a gold shell.
We find that the spaser (surface plasmon amplification by stimulated emission of radiation) conditions can be found at the poles of the scattering cross section of the system, a result that remains valid beyond the geometry studied.
We explored a wide range of parameters that cover most of the usual experimental conditions in terms of the geometry of the system and the wavelength of excitation.
We show that the conditions of spaser generation necessarily require full loss compensation, but the opposite is not necessarily true.
Our results, which are independent of the detailed response of the active medium, provide the gain needed and the wavelength of the spasers that can be produced by a particular geometry, discussing also the possibility of turning the system into optical amplifiers and SERS (surface enhanced Raman spectroscopy) substrates with huge enhancements.
We believe that our results can find numerous applications. In particular, they can be useful
for experimentalists studying similar systems in both, tuning the experimental conditions and interpreting the results.
\end{abstract}

\maketitle

\section{INTRODUCTION}
\label{sec:INTRODUCTION}

Currently there is a great interest in the interaction of plasmonic nanoparticles (NPs) and nanostructures with electromagnetic fields to optimize and increase as much as possible the magnitude of the evanescent field generated around their surfaces\cite{Zayats2005}.
These investigations are triggered by the plethora of applications arising from this property in enhanced spectroscopies such as SERS\cite{Blackie2009,Gruenke2016,Links2011,Ru2007}, TERS\cite{Beams2015,Kawata2014,Sonntag2014} (Tip Enhanced Raman Spectroscopy), able to reach the single molecule level\cite{Blackie2009,Pozzi2015,Press2012}, light focusing and imaging in the subdifraction limit of light \cite{Wu2014,Cang2015,Smolyaninov2015,Bustos-Marun2015} as well as in non linear effects such as SHG\cite{Eisenthal2006,Fan2006,Kim2008} (Second Harmonic Generation), and metamaterials with novel optical properties\cite{Kim2015,AlexandraBoltasseva2012}. In particular, it has been recently demonstrated that incorporating a gain or active media to a plasmonic nanostructure (NE) give rise to a new set of possibilities and opportunities. In this way, the optical behavior in such systems can considerably improve the performance of plasmonic devices. Theoretical studies of this phenomena could be found in different systems such as semi-shells\cite{Arnold2013}, multishells\cite{Gordon2007}, V-shape arrays\cite{Bergman2003}, core-shell nanorods\cite{Liu2011}, nanotubes and its dimers \cite{Yu2015}, nanoparticle chains\cite{Rasskazov2013,Citrin2006}, dimer emitter coupled to a metal nanoparticle\cite{Nugroho2015}, as well in metamaterials\cite{Fang2009,Hamm2011,Wuestner2011,Wuestner2010}. Also examples of experimental works on several systems can be found such as silica-core gold-shell (the system studied here)\cite{DeLuca2014}, gold-core silica-shell \cite{DeLuca2012}, gold-core sodium silicate-shell\cite{Noginov2009} , mesocapsules \cite{Infusino2014}, silver aggregates in gain media \cite{Noginov2007} or even metamaterials \cite{Zhou2013}.

On the other hand, these hybrid systems can generate new optical phenomena, such as a spaser which are nanometric sources of evanescent and propagating electromagnetic fields with a high level of wavelength tunability\cite{Stockman2009}.

From the theoretical point of view, the selection and design of suitable systems where this kind of behavior is likely to be observed constitutes a topic of paramount importance. However, most of previous studies on this topic have focused either on small systems where simple expressions can be obtained\cite{Bustos-Marun2014,Lawandy2004,Citrin2006,Nugroho2015,Rasskazov2013}, or on very demanding numerical calculations where a systematic variations of the geometry of the system is precluded \cite{Arnold2013,Bergman2003,Fang2009,Hamm2011,Li2010,Liu2011,Wuestner2011,Wuestner2010,Yu2015}.
In this respect, recently Arnold et al. studied the minimal spaser threshold for spheroidal and spherical core-shell NPs including retardation effects \cite{Arnold2016}.

In the present work we study spherical core-shell NPs where the core is made of doped-silica, providing the active medium, and the shell is made of gold, giving the plasmonic material. In this geometry one can effectively make use of the huge evanescent electromagnetic field around the plasmonic structure, for example for sensing purposes. The electromagnetic properties of this system can be calculated analytically by means of the generalized Mie's theory. This allow us to perform a deep physical analysis of the system and to relate the far and near fields in a rigorous way.
Additionally, the computation of electromagnetic properties in these structures are very fast, allowing us to explore a wide range of possible experimental conditions.
We should mention that the materials chosen as well as the dimensions of the core and the shell are feasible to be fabricated by chemical methods.\cite{Westcott1998,Sheppard2008}.
For these reasons, it is our hope that the present work could be a benchmark for experimentalist working on this or similar systems.

\section{Theory}
\label{sec:Theory}
\subsection{Active media}
\label{sec:Active media}

The interaction of photons with the conduction electrons of a plasmonic metamaterial gives rise to optical losses, which for visible light can be significant.
There are two main sources of optical losses: ohmic heating and radiative losses\cite{craig1983absorption,maier2007plasmonics}.
Several strategies have been proposed to overcome optical losses on particular examples, but probably the most promising one is the use of active media.

Active media, or gain materials, are made of dye molecules, semiconductors nanocrystals, or doped dielectrics, where there is a population inversion, created optically or electrically, that sustains the stimulated emission of radiation.
This stimulated emission is used to compensate the intrinsic optical losses of plasmonic materials. Depending on how strong is this stimulated emission of radiation compared with the optical losses, the hybrid system can be undercompensated, fully compensated or overcompensated.
These conditions can be distinguished by the value of the extinction coefficient $Q_{\mathrm{ext}}$, where $Q_{\mathrm{ext}}>0$ corresponds to undercompensation, $Q_{\mathrm{ext}}=0$ to full compensation, and $Q_{\mathrm{ext}}<0$ to overcompensation.
We adopted the term full compensation to distinguish it from $Q_{\mathrm{abs}}=0$ which correspond to the situation were only omhic losses are compensated.

If the system is undercompensated it just behaves as a regular plasmonic material but with an increased intensity of the electromagnetic fields around the plasmonic structure and narrower resonances in general.
When the system is overcompensated at frequencies far from its resonances it behaves as an optical amplifier. For frequencies close to some resonance and when losses are fully compensated, the hybrid system behaves as the nanoplasmonic counterpart of a laser, known as spaser \cite{Stockman2009}.
This behaviors can be readily understood by drawing a parallelism with conventional macroscopic lasers. The main differences are that in a spaser the optical cavity is replaced by plasmonic resonances and the electromagnetic fields are composed of propagating as well as evanescent waves.

When the system is not so close to its spaser conditions, its effect can be modeled phenomenologically on the basis of classical electrodynamics without taking into account explicitly the quantum dynamics of the ground and excited states.
This is usually done by considering the medium as a dielectric with an additional negative imaginary part added to its refractive index $n_0$, $n=n_0 + i \kappa$ (with $\kappa \leq 0$).\cite{Calander2012,Huang2015,Gordon2007,Li2010,Liu2011,Bustos-Marun2014} 
This approximation may seem too simple for the proper description of complicated systems such as active plasmonic structures. However, as soon as the dye is diluted enough in the base material that support the active media, the application of an effective medium theory \cite{craig1983absorption} to the system results in a real part of $n$ almost identical to that of the base material $\mathrm{Re}(n_0)$ and an imaginary part given by $\mathrm{Im}(n_0)+\kappa$, or in other words $n \approx n_0 + i \kappa$. Note that, although this approximation seems reasonable under the appropriate conditions, it does not strictly fulfill the Kramers-Kronig relation for the active media.\cite{craig1983absorption} Thus, our calculations should be taken in general as approximate results, useful to guide future experiments and calculations on this issue. However, when the resonances of the active medium and the spaser are close enough our results should be exact, assuming a model for the gain media as in ref. \cite{Arnold2013}.

In this work we are only taking into account the effect of the active media at precisely the frequency of excitation $\lambda_{\mathrm{exc}}$ and neglecting the effect of the real part of the refractive index of the active media. This is equivalent to consider an active media with Lorentzian wavelength dependence at resonance with $\lambda_{\mathrm{exc}}$ and with a FWHM (full width at half maximum) small enough that the spaser condition is not reached at a different $\lambda$.
The situation is analogous to use $\kappa(\lambda) =\kappa \delta(\lambda_\mathrm{exc}-\lambda) $ where $\lambda_{\mathrm{exc}}$ is the wavelength of excitation and $\delta$ is the Kronecker delta function.
Note that this is the opposite of the usual approximation of considering $\kappa$ in a wide band limit 
$\kappa(\lambda) = \kappa$ \cite{Bustos-Marun2014,Calander2012,Gordon2007,Liu2011,Li2010,Huang2015}. 
We will retake this issue later on, but we will see that, thanks to the superposition principle of linear electrodynamics, our results can be readily reinterpreted within the context of the wide band approximation or even considering more complicated wavelength dependences of $\kappa$.

There are of course some issues related to not taking into account the dynamic of the population inversion of the active media \cite{Stockman2009}.
A consequence of that appears when the system approaches the spaser condition. This results in singularities in the electromagnetic fields as well as in the extinction, aborption and scattering cross sections.\cite{Arnold2016,Baranov} This fact is indeed used as a simple numerical way of finding the spaser condition.
The behavior of the system very close to the spaser condition, i.e. true intensity of the electromagnetic fields and cross sections, is beyond the scope of the present work.

\subsection{Singularities in Mie's theory}
\label{sec:Singularities in Mie's theory}

According to Mie theory, the electric field $E_s(\vec{r})$ outside a core-shell nanoparticle (CSNP) illuminated by a plane wave is given by \cite{craig1983absorption}:
\begin{equation} \label{eq:campo}
E_s(\vec{r})=\sum_{n=1}^{\infty} E_n[i a_n N^{(3)}_{e1n}(\vec{r})- b_n M^{(3)}_{o1n}(\vec{r})] ,
\end{equation}
where the coefficients $a_n$ and $b_n$ are obtained by continuity conditions,
the vectorial functions $M_{o1m}$ and $N_{e1n}$ are the well known vector spherical harmonics of order $n$, and the the $E_n=i^n E_0 (2n+1)/n(n+1)$ functions are the projections of the plane wave in the $n$-th harmonic.
The coefficients $a_n$ and $b_n$ depend on the relative refraction indexes of the core, $m_1$, and the shell, $m_2$ (equations \ref{eq:m1}), but also on the core radius ($r$), shell thickness ($D$), and on the wavenumber $k$. Close expressions for them can be found in ref. \cite{craig1983absorption}. No surface damping corrections were performed because they are not significant for achieving a good spectral correlation with experiments as shown in ref.\cite{Hao2007} for example. The relative refractive indexes are given by:
\begin{equation}\label{eq:m1}
\begin{array}{cc}
m_1(\lambda)=\frac{n_{core}+i\kappa_{core}}{n_{med}} & m_2(\lambda)=\frac{n_{shell}+i\kappa_{shell}}{n_{med}}
\end{array}
\end{equation}
They  depend on the core ($n_{core}$ , $\kappa_{core}$) the  shell  ( $n_{shell}$ and $\kappa_{shell}$), and the non absorbing media ($n_{media}$) complex refractive indexes as well as the incident wavelength ($\lambda$)
As the refractive indexes are assumed to be only a function of $\lambda$, they are unambiguously specified by the illumination wavelength $\lambda$.
The active media is given by a doped silica. We will assume that its real part is the same than in the pure bulk material, unchanged by the presence of the dopant, and only its imaginary part ($\kappa_{core}$) change with the gain strength. Therefore, once we set which materials are the core and shell of the NE, there are only four independent variables ($r$, $D$, $\lambda$ and $\kappa_{core}$) to be evaluated, to systematically investigate gain effects at different geometries and illumination conditions. For simplicity in the following we will refer to $\kappa_{core}$ as just $\kappa$.

\subsubsection{Poles of $Q_{sca}$ and $\Gamma(\vec{r})$}
\label{sec:Poles of Q_sca and Gamma}

For a given core shell nanostructure, its  complex  field (eq \ref{eq:campo}) at each point of the space $\vec{r}$,  and the corresponding scattering cross section  (eq \ref{eq:csca} below) are described in terms of the same set of coefficients, $a_n$  and $b_n$.
We are interested in the magnitude of the near electric field enhancement, $\left|\Gamma(\vec{r})\right|^2 =\left| E_s(\vec{r}) / E_0 \right|^2 $, but as will become clear soon there is a direct correspondence between $C_{sca}$ and  $|\Gamma|^2(\vec{r})$, as both quantities depends on the the square modulus of $a_n$ and $b_n$.
\begin{equation}\label{eq:csca}
C_{sca}=\frac{2\pi}{k^{2}}\sum_{n=1}^{\infty} (2n+1)(|a_n|^2 +|b_n|^2) 
\end{equation}
Even though we are not giving the expressions for rational functions $a_n$ and $b_n$ it is enough to mention that for a given set of values of the independent variables ($r$, $D$, $\lambda$ and $\kappa$), there is a condition under which the expression in the denominators of $a_n$ and $b_n$ vanishes totally (real and imaginary parts).
This condition leads to a divergency, or a pole, of $C_{sca}$ as well as $\Gamma(\vec{r})$.
As we approach a divergency, one of the multipoles dominates the expansions in eqs. \ref{eq:csca} and \ref{eq:campo}, therefore the electric field $E_s$ and the scattering cross section ($C_{sca}$) can be approximated by:
\begin{equation}\label{eq:aprox-campo}
E_s(\vec{r})\simeq\ i^{n+1} E_0 a_n\frac{2n+1}{n(n+1)}N^{(3)}_{e1n}(\vec{r}),
\end{equation}
and
\begin{align} \label{eq:aprox-sca}
C_{sca}&=\pi (r+D)^2 Q_{sca} \simeq\frac{2\pi}{k^{2}} (2n+1)|a_n|^2 \notag \\ 
&=\frac{\lambda^{2}}{2\pi n_{med}^{2}} (2n+1)|a_n|^2 ,
\end{align}
where $ Q_ {sca} $ is the scattering efficiency (the ratio between $ C_{sca} $ and the geometric area of the nanostructure). Taking the square modulus of eq \ref{eq:aprox-campo}, and considering  eq \ref{eq:aprox-sca}, one can readily 
obtain an analytical expression for the squared field enhancement, $\Gamma ^ 2(\vec {r})$, at each point outside the NE,
\begin{align}\label{eq:incremento}
|\Gamma(\vec{r})|^2 &\simeq  \frac{\pi^2 (r_{core}+D)^2}{\lambda^2 n_{med}^{2}} \frac{(4n+2)}{(n(n+1))^2} \notag \\
& \times Q_{sca} \left|N_{e1n}^{(3)}(\vec{r})\right|^2
\end{align}
This result is general and valid for any system of concentric spheres in the surroundings of their poles or singularities. 
For nonspherical nanoparticle one can still expand the fields in eigenmodes and obtain equations equivalent to eqs. \ref{eq:campo}, \ref{eq:csca} and \ref{eq:incremento}. In this case, eq. \ref{eq:incremento} will have a different prefactors multiplying $Q_{sca}$ and different eigenfunctions.
However, the connection among $|\Gamma|^2$ and $Q_{\mathrm{sca}}$ will still be valid close to a pole.

Eq. \ref{eq:incremento} indicates that finding the poles of the scattering efficiency is equivalent to search for the poles of the field enhancement. This has the advantage that one does not need to care about the specific regions in space outside the CSNP where the enhancement occurs.

The physical origin of these divergences are interpreted as the spaser conditions, which implies full loss compensation at a resonant frequency. In the next section we will see that singularities are always located at the full loss compensation condition.

\subsubsection{Poles of $Q_{ext}$ and $Q_{abs}$}
\label{sec:polos of qext y qabs}

Now, let us consider the expressions for $C_{ext}$ and $C_{abs}$: 
\begin{equation} \label{eq:cext}
C_{ext}=\frac{2\pi}{k^{2}}\sum_{n=1}^{\infty} (2n+1)Re[a_n +b_n] 
\end{equation}
\begin{align}\label{eq:abs}
C_{abs} & =C_{ext}-C_{sca} \notag \\
& =\frac{2\pi}{k^{2}}\sum_{n=1}^{\infty} (2n+1)(Re[a_n +b_n]-|a_n|^2-|b_n|^2)
\end{align}
As it has been shown for $C_{sca}$, there are sets of values of $r_ {core}$, $D$, $\lambda$ and $\kappa$ that make zero or almost zero the denominator of a particular term $a_n$ (or $b_n$) which then dominates the summation.
Taking this approach we can write this dominant coefficient as the ratio of two complex functions $f=f'+i f''$ and $g=g'+i g''$, where we denote the real and imaginary parts as $f'$, $g'$ and $f''$, $g''$ respectively, i.e. $a_n={f_{a_n}}\div{g_{a_n}} $ (or $b_n={f_{b_n}}\div{g_{b_n}} $). Then eqs. \ref{eq:cext} and \ref{eq:abs} can be rewritten as:
\begin{equation}\label{eq:aprox-ext}
C_{ext}\simeq\frac{2\pi}{k^{2}}(2n+1)\frac {f'g'+f''g''}{|g|^2} 
\end{equation}
\begin{align} \label{eq:aprox-abs}
C_{abs} \simeq \frac{2\pi}{k^{2}}(2n+&1)\left(\frac {f'g'+f''g''} {|g|^2}  -|a_n|^2\right) 
\end{align}

Eq. \ref{eq:aprox-ext} presents a divergence in the limit when $g$ vanishes and therefore the absolute value of $C_{ext}$ goes to infinity, although its sign changes at this point. For a fixed geometry at resonant wavelength, the change of sign of $C_{ext}$ can be readily understood if we expand the function $g$ around $\kappa_{pole}$, where $g(\kappa_{pole})=0$, $g(\kappa) \approx \left . \frac {dg}{d \kappa} \right|_{\kappa_{pole}} (\kappa-\kappa_{pole})$. Then, the change of sign of $g$, and thus of $C_{ext}$, requires only a nonzero $dg/d\kappa$ which is our case.
This feature is interesting because a change of sign of $C_{ext}$ matches the full loss compensation condition. This implies that, similarly to that found in ref. \cite{Stockman2010}, the conditions of spaser generation necessarily requires full loss compensation in our system, even beyond the quasi-static limit.

$C_{ext}$ is the total power deflected from the incident plane wave by scattering and absorption, thus its negative value implies an overall energy release outside the particle. We call this an optical amplifier due to its potential applications for information transport at the nanoscale. The value of $\kappa$ at which this occurs will be called $\kappa_{flc}$, where $flc$ stands for full loss compensation.
As discussed above, poles should always fall over the $\kappa_{flc}$ curves, but due to the discreteness of the pole's conditions in finite systems, full loss compensation not necessarily implies a pole. In the case of periodic infinite systems, where discrete resonances turn into bands, it is still possible to have full loss compensation without reaching the spaser condition, outside the bands.

Around a pole, the last term of eq \ref{eq:aprox-abs} (which comes from $C_{sca}$) dominates. As a consequence, $C_{abs} \approx -C_{sca}$. Then the singularity of $|\Gamma|^2$ can be found either as a positive singularity of $C_{sca}$ or a negative singularity of $C_{abs}$. The other alternative, using $C_{ext}$, is also possible but cumbersome to apply in practice due to the change of sign.

\section{Results and discussion}
\label{sec:Results and discussion}

\subsection{System studied}

\begin{figure}
\includegraphics[width=1.7in]{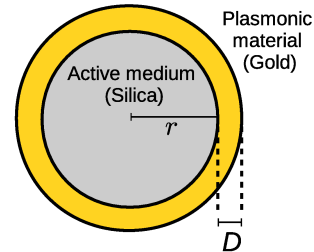}
\caption{System studied.}
\label{fig:system}
\end{figure}

The system studied (shown in fig. \ref{fig:system}) consists of a core and shell of radius $r$ and thickness $D$ respectively, where the core is made of silica and contains the appropriate dopant while the shell is made of gold.
This system was chosen mainly by its experimental feasibility
and the possibility to control precisely the core and shell geometries\cite{Oldenburg1998,Razink2007,Jana2001}.

The real part of the refractive index of the silica core, $n_0$, was taken from ref. \cite{Malitson1965}. As in many previous works,\cite{Calander2012,Huang2015,Gordon2007,Li2010}, its gain character is emulated by adding a negative imaginary part to $n_0$, $n=n_0+i \kappa$, where $\kappa \leq 0$. The complex refractive index of gold is given by a cubic interpolation of the experimental data of ref. \cite{PhysRevB.6.4370} The particle was assumed to be immersed in an aqueous medium with a refractive index of 1.33.
$Q_{sca}$, $Q_{ext}$ and $Q_{abs}$,  were calculated by using standard Mie theory.
The values of $\kappa$ that make $Q_{ext}=0$, i.e full loss compensation condition, will be denoted as 
$\kappa_{flc}$, while the values of $\kappa$ that make $Q_{abs}=0$, ohmic loss compensation condition, will be denoted as $\kappa_{olc}$.
As mentioned , in our calculation we are only considering the effect of $\kappa$ at the wavelength of excitation, which is equivalent to consider $\kappa(\lambda) = 0$ for $\lambda$s different from the wavelength of observation and/or excitation.

\subsection{General behavior}
\label{sec:General behavior}

\begin{figure*} 
 \begin{center}
  \subfigure{\includegraphics[width=3.2in,trim=0.0in 0.0in 0.0in 0.0in, clip=true]{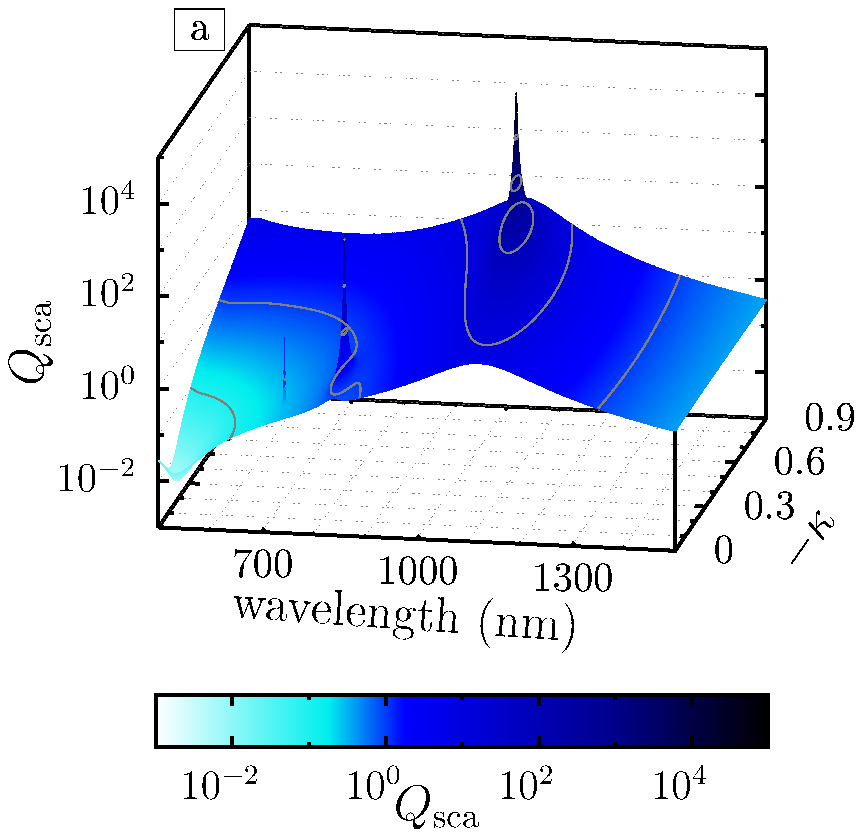}}
  \subfigure{\includegraphics[width=3.2in,trim=0.0in 0.0in 0.0in 0.0in, clip=true]{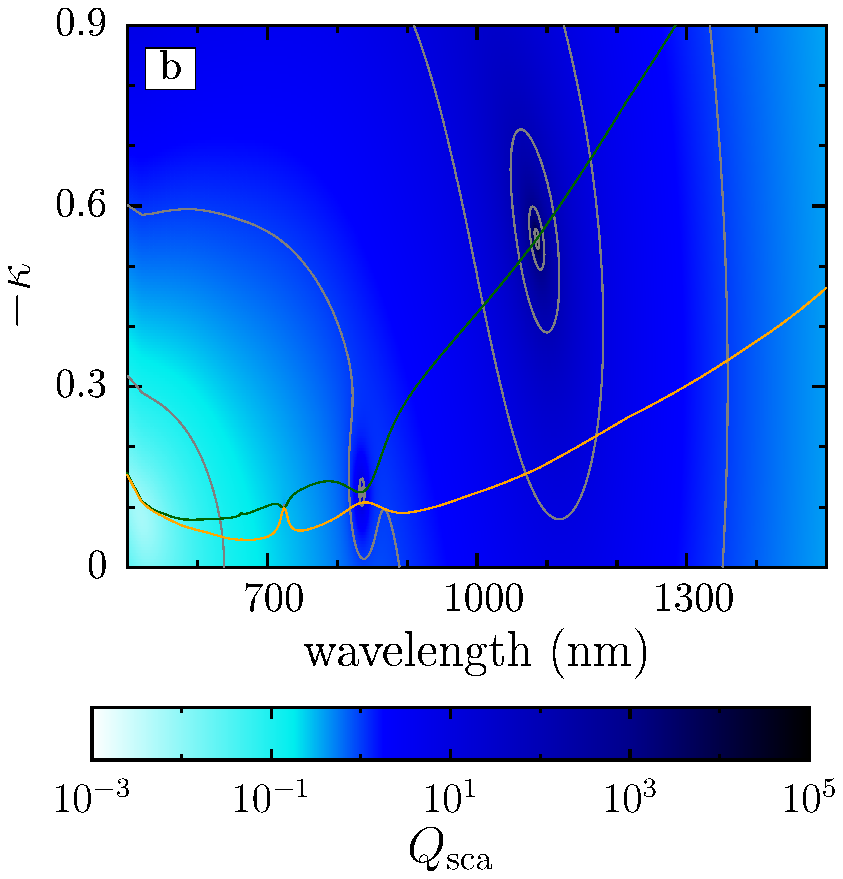}} \\
  \subfigure{\includegraphics[width=3.2in,trim=0.0in 0.0in 0.0in 0.0in, clip=true]{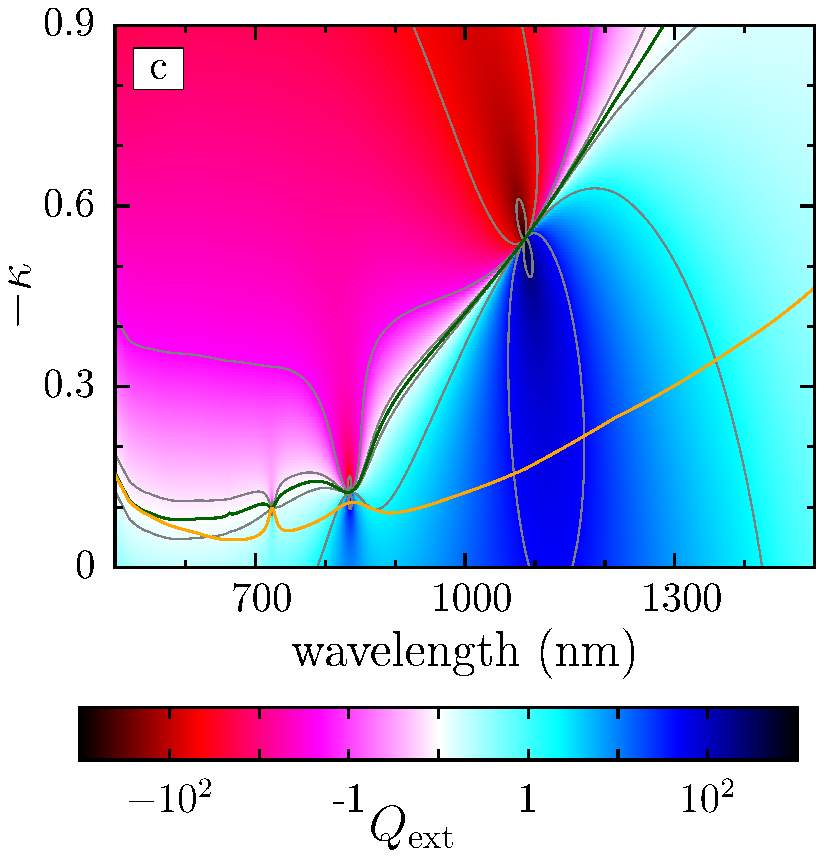}}
  \subfigure{\includegraphics[width=3.2in,trim=0.0in 0.0in 0.0in 0.0in, clip=true]{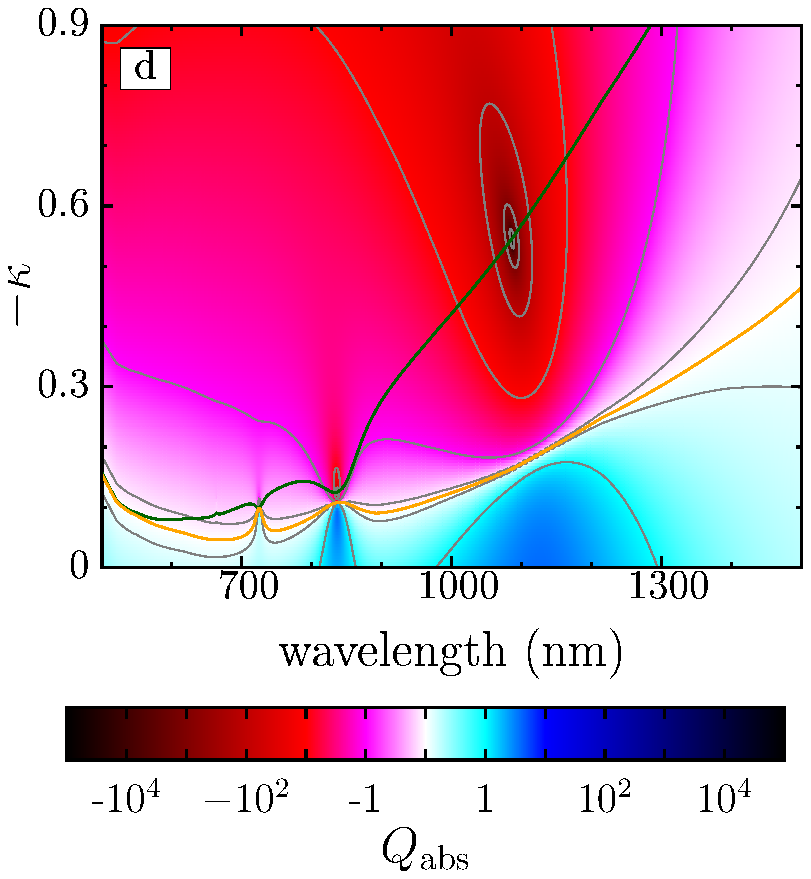}}
 \end{center}
 \caption{Scattering cross sections of a gold-coated active NP of $r=75nm$ and $D=5nm$.
 \textbf{a)} 3D-plot of $Q_{sca}$ vs $\kappa$ and $\lambda$. \textbf{b)} Contour plot of the left figure. \textbf{c)} contour plot of $Q_{ext}$ vs $\kappa$ and $\lambda$. \textbf{d)} contour plot of $Q_{abs}$ vs $\kappa$ and $\lambda$. The color scale is logarithmic except panels \textbf{c)} and \textbf{d)} between -1 and 1, where it is linear.
 Green continuous line marks $\kappa_ {flc}$, where ``$flc$'' stands for full loss compensation condition. Orange line marks $\kappa_ {olc}$, where ``$olc$'' stands for Ohmic loss compensation condition. Gray lines marks the contours of the positive and negative powers of ten starting from $10^{-1}$.}
 \label{fig:Qs}
\end{figure*}

In order to understand the problem of finding and describing the conditions under which singularities are encountered, we will start with a representative fixed geometry.
For no special reason, we selected $r_{core} =75 nm$ and $ D=5 nm$ for this purpose.
For this geometry we performed a systematic study of the variation of $Q_ {sca}$, $Q_ {ext}$, and $Q_ {abs}$ in the space generated by the remaining variables, $\lambda$ and $\kappa$.
The values of $Q_{sca}$ obtained as a function of $\kappa$ and $\lambda$ are displayed, in logarithmic scale, in the the 3D plot shown in fig. \ref{fig:Qs} a). There, the poles discussed in section \ref{sec:Singularities in Mie's theory} can be clearly seen. For the sake of  a more comprehensive analysis, fig. \ref{fig:Qs} also depicts a contour plot of this 3D figure, in panel b). In this panel, the green and orange continuous lines represent the values of $\kappa$ in which $Q_{ext}$ and $Q_{abs}$ present a change on their signs, $\kappa_{flc}$ and $\kappa_{olc}$ respectively.
Similarly to that found by Stockman in ref. \cite{Stockman2010} in the quasi-static limit, here we can see that the spaser condition, given by the divergences of $Q_{sca}$, implies always the full loss compensation condition. Note that the opposite is not necessarily true, due to the discreteness of the resonances. Three poles or singularities are easily distinguished in this figure, they are located at: ($\lambda = 1085.9 nm$, $\kappa=-0.5450$) ; ($\lambda = 835.6 nm$, $\kappa = -0.1258$) and  ($\lambda = 724,5 nm$, $\kappa= -0.0923$).
Panels c) and d) of figure \ref{fig:Qs}  show $Q_{ext}$ and $Q_{abs}$ projected for the same range of $\lambda$ and $\kappa$. In these panels also the continuous and dotted red lines correspond to $\kappa_{flc}$ and $\kappa_{olc}$ respectively.
\begin{figure*}
 \begin{center}
 \subfigure{\includegraphics[width=2.1in,trim=0.1in 0.in 0.in 0.0in, clip=true]{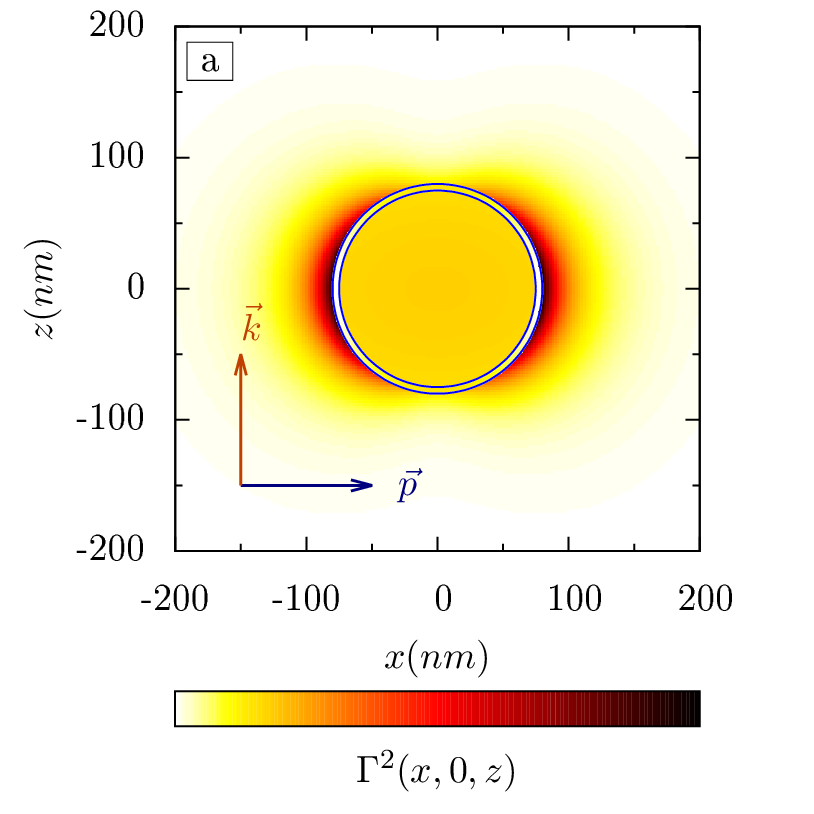}}        
 \subfigure{\includegraphics[width=2.1in,trim=0.1in 0.in 0.in 0.0in, clip=true]{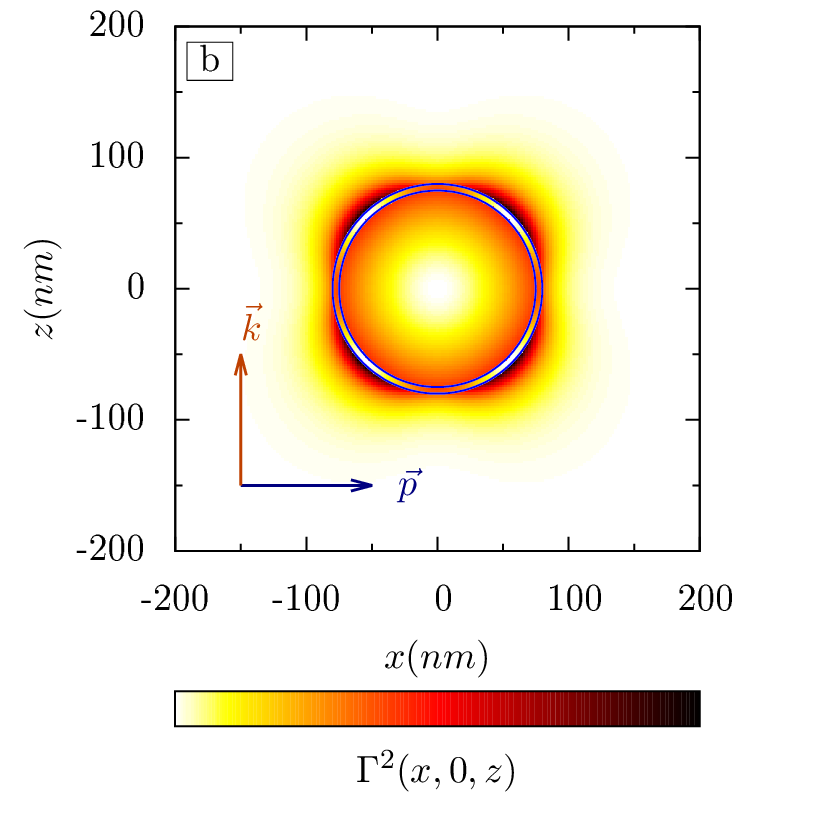}}
  \subfigure{\includegraphics[width=2.1in,trim=0.1in 0.in 0.in 0.0in, clip=true]{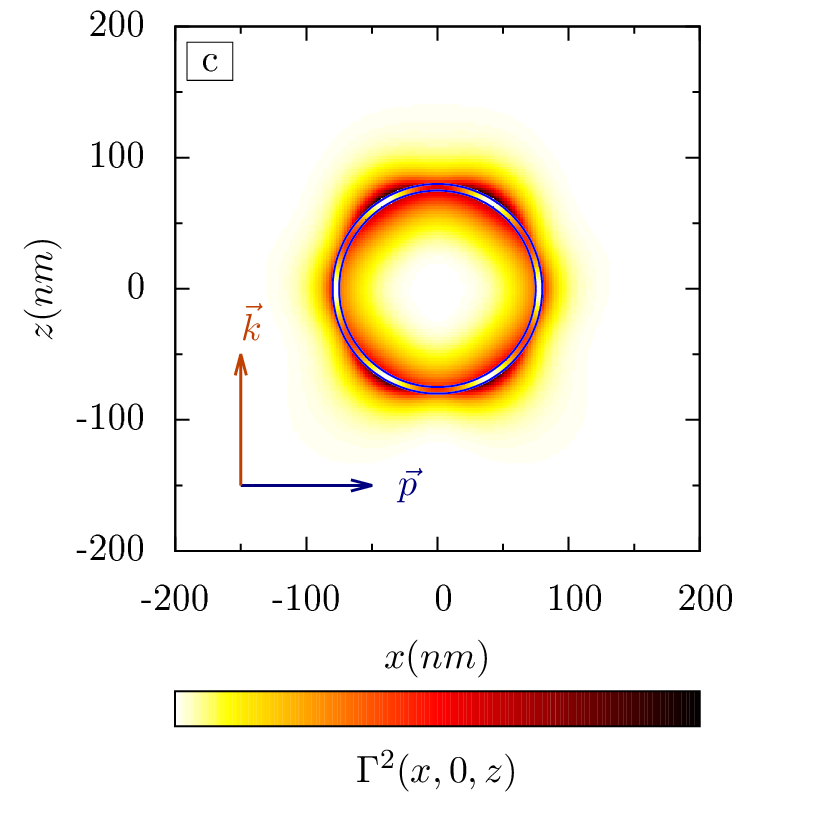}}
 \end{center}
 \caption{Electric field enhancement in the $xz$ plane, $\Gamma^2(x,0,z)$, for a core-shell NP with $r=75nm$ and $D=5nm$ at three different  $\lambda$ and $\kappa$  (near the three poles encountered). $\Gamma^2$ is shown in arbitrary scale. The orange and blue arrows represents the directions of propagation $\vec{k}$ and polarization $\vec{p}$ respectively. \textbf{a)}  $\lambda = 1085.9 nm$ and $\kappa = -0.5450$. \textbf{b)}  $\lambda = 835.6 nm $ and $\kappa = -0.1258$.\textbf{c)}  $\lambda = 724.5 nm $ and $\kappa = -0.0923$}
 \label{fig:75@80}
\end{figure*}

The orange and green lines define three distinctive regions corresponding to three different regimes of energy losses.
For small $\kappa$, between the bottom axis and the orange line, the values of $Q_{ext}$, $Q_{abs}$ are both positive indicating that the radiative and dissipative energy losses are not compensated.
For intermediate values of $\kappa$, between the orange line and the green line, $Q_{abs}$ is negative while $Q_{ext}$ is still positive (figs \ref{fig:Qs} c) and d)). This implies that the gain  media is able to compensate dissipative losses but not radiative ones. Under these conditions, more radiation is going out of the system than the incident one. However the system still looks under-compensated. The intensity of the forward radiation is lower than the incident one.
For large values of $\kappa$, beyond the continuous green line, $Q_{ext}$ is negative indicating that the active medium is able to fully compensate both radiative and dissipative losses. There, the system is in principle able to amplify the incoming radiation producing an outgoing wave of the same or more intensity than the incoming one. This region must be taken with caution because if this occurs at the wavelength of resonance ($\lambda = \lambda_{pole}$), the system will start to increase its energy with time. In the present formalism, as we are not taking into account the dynamic of the excited and ground states of the dye molecules or the dopants, this will cause the divergence of the electromagnetic fields that allows us to identify the spaser condition.

Fig. \ref{fig:75@80} shows the near field at conditions very close to the first three poles shown on fig \ref{fig:Qs}. 
For each pole, the near field enhancement was calculated with the BHFIELD program \cite{Suzuki2008}. We performed a exploration of the $xz$ plane through the middle of the nanoparticle with polarization and propagation along the $x$  and $z$ axis respectively within a square of side 400 nm with 2 nm wide grid.
The first pole, which corresponds to the highest $\lambda$, clearly can be assigned to a dipole mode while the second one corresponds to a quadrupole and the third to an octupole.

Finally we want to mention that, as fig. \ref{fig:Qs} shows, the poles are always blue shifted with respect to the maximum at $\kappa=0$. It is important to consider this point when studying plasmonic active systems, as one could make the mistake of trying to find the lasing condition at the wavelength of the maximum at $\kappa=0$.
The blue shift of the poles as $\kappa$ increases, is expected recalling the behavior of damped harmonic oscillators. There, introducing energy losses give rise to a red shift of the resonant frequencies, along with a spectral broadening and a decrease of the peak intensities.  Consequently, increasing the gain, which decreasing the damping, should produce a blue shift of the peaks together with a narrowing of their width and an increment of their height. 

\subsection{Effect of the wavelength dependence of the gain}
\label{sec:Effect of a wavelength dependence of the gain}

\begin{figure} 
 \begin{center}
  \subfigure{\includegraphics[width=3.2in,trim=0.0in 0.02in 0.0in 0.17in, clip=true]{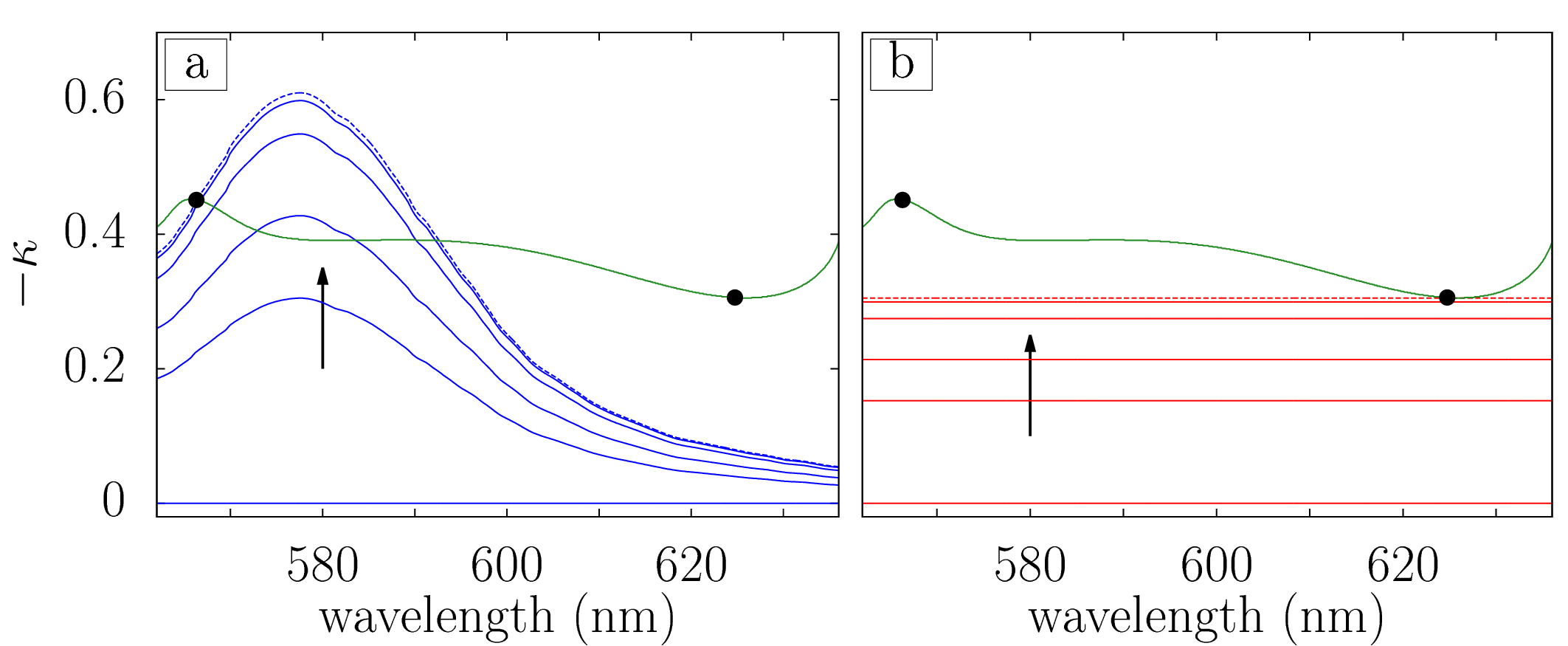}} \\
  \subfigure{\includegraphics[width=3.2in,trim=0.0in 0.02in 0.0in 0.17in, clip=true]{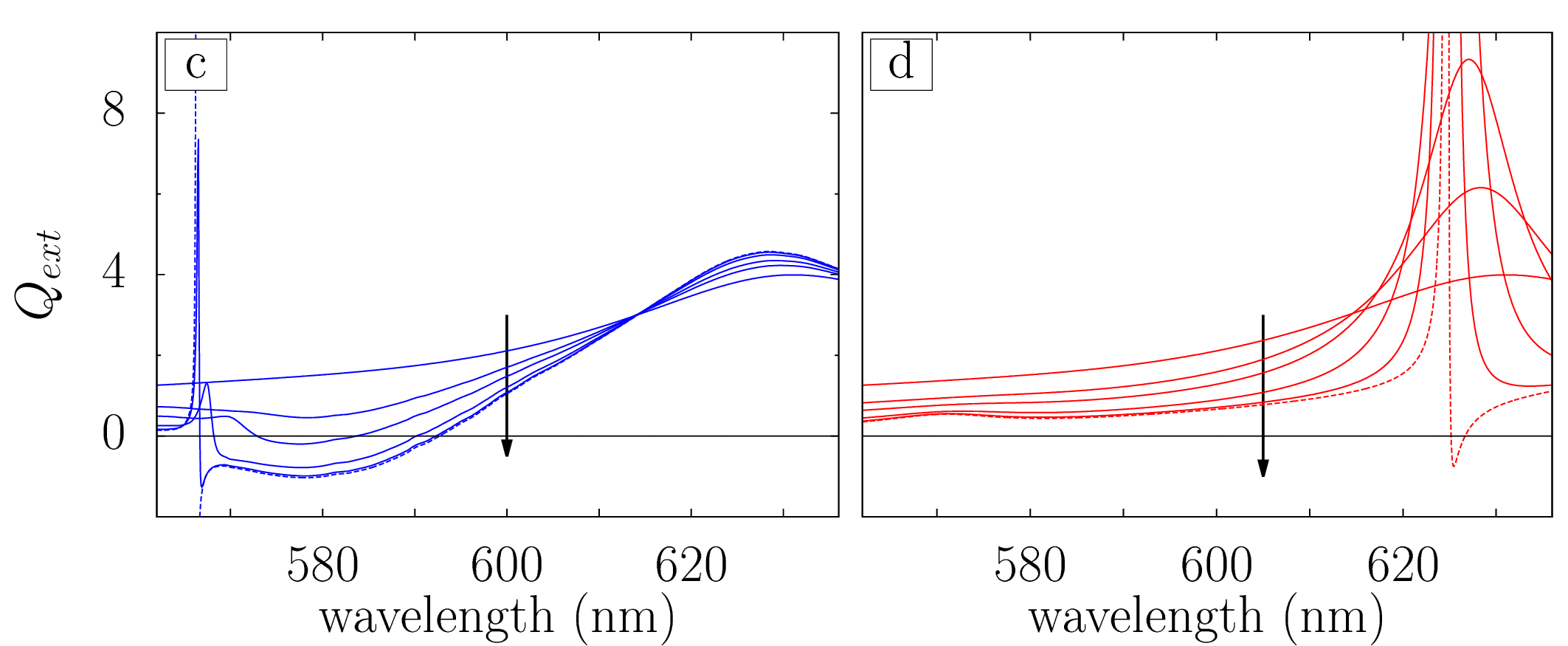}} \\
  \subfigure{\includegraphics[width=3.2in,trim=0.0in 0.02in 0.0in 0.17in, clip=true]{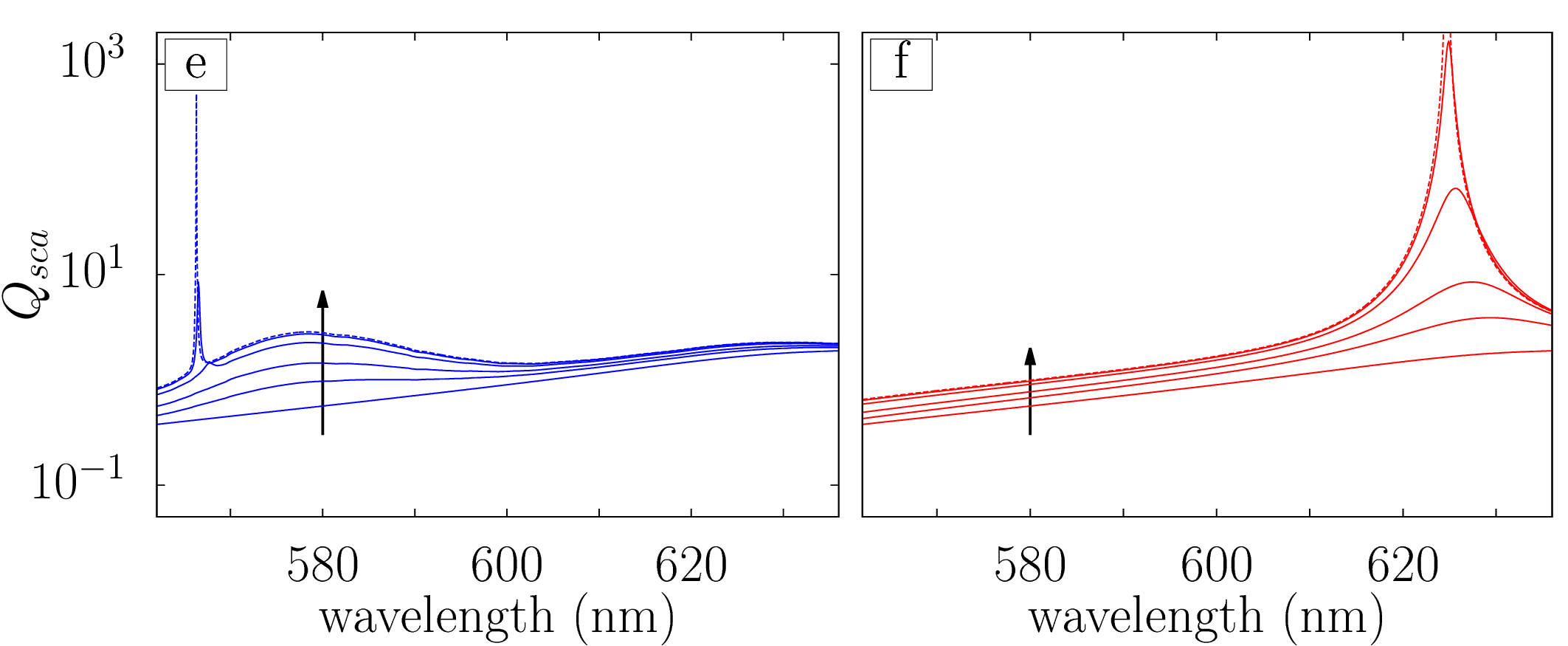}}
 \end{center}
\caption{Top panels (a and b) show the wavelength dependence of $\kappa$ for different strengths of the active medium, in increasing order $\kappa/\kappa_{\mathrm{pole}}=0.0$, 0.5, 0.7, 0.9, 0.98, and 0.99. Left panels (a, c, and e) used a realistic wavelength dependence assuming the dye is rhodamine B, whose emission spectrum was taken form ref. \cite{radhamine}. Right panels (b, d, and f) assume a wide band approximation. Full loss compensation, $Q_{\mathrm{ext}}(\lambda, \kappa)=0$, is indicated by a green continuous line while the spaser conditions ($\lambda_{\mathrm{pole}},\kappa_{\mathrm{pole}}$) are indicated by black circles. Middle (c and d) and bottom (e and f) panels show the extinction and scattering coefficients respectively calculated with the values of $\kappa(\lambda)$ shown in the top panels. The direction of the arrows indicate increasingly higher values of $\kappa$.  We used $r=65$nm and $D=12$nm.}
\label{fig:example}
\end{figure}

Up to this point, our treatment has neglected the wavelength dependence of active media, which in our case was equivalent to consider $\kappa(\lambda)$ as a Kronecker delta function at the wavelength of excitation. This allowed us to keep our analysis independent of the particularities of the active medium and to focus only on its general effects.
However, the problem of adding a $\lambda$ dependence of $\kappa$ can be readily treated.
Essentially, due to the principle of superposition, the behavior of systems with frequency dependent gains can be obtained by the appropriate weighting of our results.
Fig. \ref{fig:example} shows two examples of wavelength dependences of $\kappa$. Left panels (a, c, and e) assume the dye is rodhamine B while right panels (b, d, and f) assume a flat dependence, or a wide band approximation.
The values of $Q_{\mathrm{ext}}$ and $Q_{\mathrm{sca}}$ at each frequency were calculated using the values of $\kappa(\lambda)$ shown in the top panels (a and b). Two things are interesting to note in the figure. First, the wide band approximation should be used with care when two poles are not close enough. As shown in the example, blindly using this approximation can lead to wrong estimations of the observed $\kappa_{\mathrm{pole}}$ and $\lambda_{\mathrm{pole}}$.
Second, the non flat dependence of $\kappa$ on $\lambda$ is what enables the system to go from a regular plasmonic structure to a spaser by passing first through an optical amplifier behavior. A gradual increase of the gain strength without a $\lambda$ dependence will preclude the observation of the optical amplifier behavior.

\subsection{Effect of the geometry of the system}
\label{sec:Effect of the geometry of the system}

In this section we discuss the dependence of the positions of the different poles with the geometric parameters that describe the morphology of the system: $r$ and $D$. This analysis may result specially useful for experimentalists, as it gives the conditions that will produce active-nanoplasmonics systems with lasing activity in a desired wavelength.

In order to find the poles for a given geometry, we first made numerical calculations of $Q_{sca}$ as a function of $\lambda$ and $\kappa$, similarly to section \ref{sec:General behavior}. All local maximum in $Q_{sca}$ were recorded. 
In order to distinguish true divergences from simple maxima, we performed a simplex\cite{Press:2007:NRE:1403886} optimization starting from each maximum, using $Q_{sca}^{-1}$ as the cost function and a quadratic interpolation of the value of the shell refractive index.
We repeat this procedure varying systematically $r$ and $D$. The parameter $r$ was  varied from 50nm to 150nm at steps of 5nm and the parameter $D$ was changed from 5nm to 30nm at steps of 1nm. The intervals in $\lambda$ and $|\kappa|$ taken were 350-1500nm and 0-3 respectively

\begin{figure*}[tbph]
  \includegraphics[width=5in]{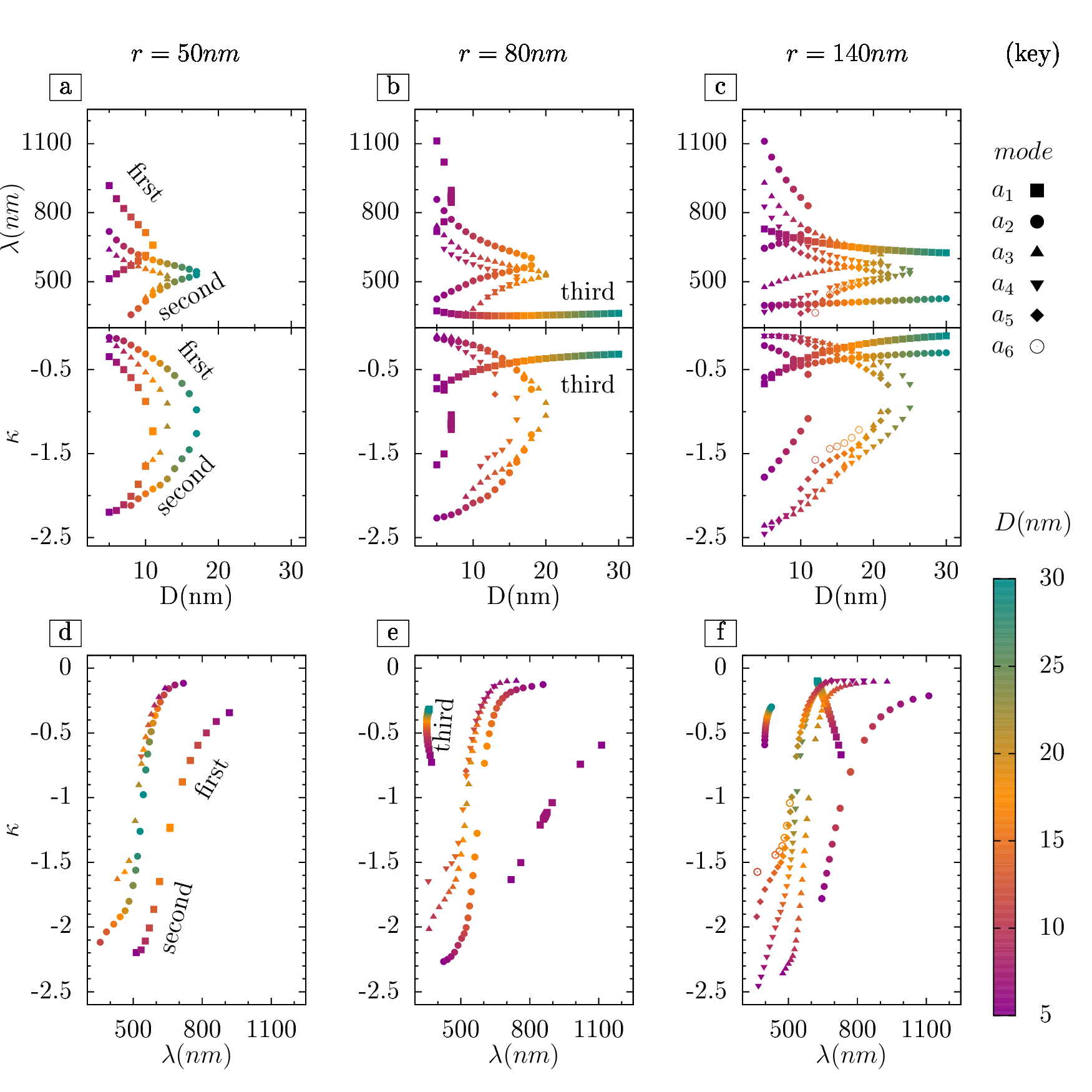}
   \caption{Poles positions of gold-coated active NPs of three different radius (in columns).
  The first row shows the dependence of $\lambda$ coordinate of the pole with the geometrical parameter $D$. The second row shows the dependence of $\kappa$ coordinate of the pole with $D$. The third row summarizes the two previous dependences on $\kappa$ and $\lambda$ in a single plot. All graphics share the same color key and the same symbols, which depend on $D$ and the mode respectively.}
    \label{fig:Rbigs}
\end{figure*}

Close to a pole, two conditions should be fulfilled. First, the contribution of the dominant mode to $Q_{sca}$ should approach 1, and second the value of $Q_{sca}$ should go to infinity.
We considered that a set of values $\kappa_{pole}$ and $\lambda_{pole}$ corresponds to a true pole when $Q_{sca}>10^4$ and the contribution of the dominant mode to $Q_{sca}$ was higher than $0.999999$.
Some representative examples of $\kappa_{pole}$ and $\lambda_{pole}$ versus $D$ (for fixed $r$) are shown in fig. \ref{fig:Rbigs}. The complete calculations are provided as supporting information (SI).

We should mention, that the methodology described above should be equivalent to directly finding the zeros of the denominators of the $a_n$ coefficients as in ref. \cite{Calander2012, Huang2015}. However, this last method requires the knowledge of the analytical expressions for $a_n$. Our proposal, instead, could be readily adapted to almost any currently used method for the numerical calculation of scattering cross sections. Therefore, it can be applied to NPs of arbitrary shape.

Fig. \ref{fig:Rbigs} shows the typical behavior of varying $D$ for small, medium and large cores ($r=50, 80, 140 nm$).
In the top panels (a, b, and c) it can be observed the lasing conditions in terms of $\lambda$ and $\kappa$ separately.
The bottom panels (d, e, and f) summarize the information sown in the above panels (a, b, and c) by showing the values $D$ as a function of both $\kappa$ and $\lambda$.
The value of $D$ is encoded as a color index and the mode that produce the lasing can be distinguished as different dot types.
The dominant mode indicated in the figures was labeled with the name of the coefficients of eq. \ref{eq:campo}.
Note that several poles can correspond to the same $a_n$ coefficient. This occurs just because a pole is a zero in the denominator of the coefficients of eq. \ref{eq:campo} and this condition can in principle be fulfilled for several pairs of values of $\lambda$ and $\kappa$, depending on the dielectric constant of the materials involved. All this poles will share the same electromagnetic field profile outside the NP, as shown in eq. \ref{eq:incremento}, but will differ in the distribution of the electromagnetic field inside the NP. The inner fields are not relevant for sensing applications and generally not accessible experimentally. Therefore, for the purposes of the present work we will skip the discussion regarding this issue.

For small cores the systems behave as shown in the left column of fig. \ref{fig:Rbigs}.
Typically two branches of the same mode and opposite $\lambda$ (or $\kappa$) dependence on $D$ are observed. These branches are denoted as ``first'' and ``second'' in this figure. 
Note that there is a critical value for which the poles collapse into exceptional points.\cite{Rotter} This phenomenon has been previously reported in other active plasmonic systems.\cite{Bustos-Marun2014}
In the lower panel of the left column of fig. \ref{fig:Rbigs}, one can notice that, for both branches, $|\kappa_{pole}|$ decreases as $\lambda_{pole}$ increases.
This behavior is in agreement with ref. \cite{Huang2015} and it is consistent with the dependence of the intrinsic losses of the materials with $\lambda$, larger $\lambda$s correspond to smaller losses. For small cores, this effect always overcome the increase in the total amount of absorbing material with $D$. This can be clearly seen in the case of the second branches, where increasing $D$ gives anyway a smaller $|\kappa_{pole}|$ (see panel d).

For intermediate cores, central column of fig. \ref{fig:Rbigs}, modes of higher order appear but they follow the same $\lambda$ vs $D$ (or $\kappa$ vs $D$) trend discussed above.
The difference is the appearance of a third pole corresponding also to a $a_1$ mode, labeled in fig. \ref{fig:Rbigs} as ``third''.
As mentioned, the number of poles result from equating the denominator of the $a_n$ coefficients to zero which can gives more than two solutions. The appearance of a third pole has been recently reported, even for small particles in the quasi-static limit \cite{Arnold2016}. Probably only due to numbers, we did not find this third solution for particles with small cores in the range of parameters studied.
Notice that this extra mode does not follow the trend described for the first and second branches, that is, increasing $\lambda_{pole}$ is not correlated with smaller $|\kappa_{pole}|$.
Lets recall that poles labeled with the same $a_n$ index should present the same electromagnetic field profile outside the NP.
Then, at a given radius and shell thickness, the radiative losses must be the same for all poles corresponding to the same order.
Their different $\kappa_{\mathrm{pole}}$ values must arise form differences in the ohmic losses, which we are not able to explain systematically for the third pole.
Note that this pole not always follows the same $\kappa_{pole}$ vs $\lambda_{pole}$ trend, see panels e and f.
For larger cores (c and f panels) the main difference is the appearance of additional higher order modes as expected.
In general as $r$ increases, and for the same multipolar order, there is a shift of the poles to smaller $D$ values. For a given multipole, this is accompanied by a red shift of $\lambda_{\mathrm{pole}}$ with $r$.

\section{Conclusions}
\label{sec:conclusions}
We have thoroughly studied the wavelength and gain dependence of the response of core-shell nanospheres made of silica and gold, and where the silica core acts as an optically active medium.
The system studied is feasible experimentally and has the advantage that hot spots around the plasmonic structure are in principle physically accessible for sensing purposes.
We have demonstrated analytically that for CSNP with gain, the magnitude of the field enhancement is proportional to the scattering cross section.
We have used this result to find the spaser conditions directly from the poles of the scattering cross sections.
As discussed in section \ref{sec:Singularities in Mie's theory}, this method can also be applied to NPs of other shapes.
We have found that the spaser conditions always fall on the curves given by $\kappa_{flc}$ vs $\lambda$, where $\kappa_{flc}$ stands for the value of $\kappa$ at which full loss compensation condition is reached.
The curves of $\kappa_{flc}$ vs $\lambda$ also determine the condition for optical amplification which can be troublesome for this last application.
However, due to the discreteness of the spaser conditions, it is possible to tune the system to act as an spaser or as an optical amplifier, provided the frequency response of the active medium is narrow enough.
We report the different spasing conditions for each multipolar mode, available for the set of geometrical parameters that define the morphology of the system.
Our systematic study have covered a wide range of possible experimental conditions, which can result especially useful for experimentalists working on similar systems.
We believe our results will be useful for many applications, including optical amplification, but especially for sensing as it is known that the near fields produced by spasers are huge, even higher than those of normal metallic NPs.

\section{Acknowledgements}
\label{sec:acknowledgements}
The authors acknowledge the financial support from CONICET, SeCyT-UNC, ANPCyT, and MinCyT-Cordoba.

\section{Associated content}
Supporting Information Available: Full list of values of the core radius Rcore, shell thickness D, Pole wavelength $\lambda$ and the imaginary part of the active media refractive index $\kappa$ and the nature of the mode $a_n$ that give rise to the spaser conditions of spherical nanoparticles made of an active silica core and a gold shell.
Supporting Information available at: http://pubs.acs.org/doi/abs/10.1021/acs.jpcc.6b05240.

\bibliography{./Divergencies13.bib}

\begin{thebibliography}{60}%
\makeatletter
\providecommand \@ifxundefined [1]{%
 \@ifx{#1\undefined}
}%
\providecommand \@ifnum [1]{%
 \ifnum #1\expandafter \@firstoftwo
 \else \expandafter \@secondoftwo
 \fi
}%
\providecommand \@ifx [1]{%
 \ifx #1\expandafter \@firstoftwo
 \else \expandafter \@secondoftwo
 \fi
}%
\providecommand \natexlab [1]{#1}%
\providecommand \enquote  [1]{``#1''}%
\providecommand \bibnamefont  [1]{#1}%
\providecommand \bibfnamefont [1]{#1}%
\providecommand \citenamefont [1]{#1}%
\providecommand \href@noop [0]{\@secondoftwo}%
\providecommand \href [0]{\begingroup \@sanitize@url \@href}%
\providecommand \@href[1]{\@@startlink{#1}\@@href}%
\providecommand \@@href[1]{\endgroup#1\@@endlink}%
\providecommand \@sanitize@url [0]{\catcode `\\12\catcode `\$12\catcode
  `\&12\catcode `\#12\catcode `\^12\catcode `\_12\catcode `\%12\relax}%
\providecommand \@@startlink[1]{}%
\providecommand \@@endlink[0]{}%
\providecommand \url  [0]{\begingroup\@sanitize@url \@url }%
\providecommand \@url [1]{\endgroup\@href {#1}{\urlprefix }}%
\providecommand \urlprefix  [0]{URL }%
\providecommand \Eprint [0]{\href }%
\providecommand \doibase [0]{http://dx.doi.org/}%
\providecommand \selectlanguage [0]{\@gobble}%
\providecommand \bibinfo  [0]{\@secondoftwo}%
\providecommand \bibfield  [0]{\@secondoftwo}%
\providecommand \translation [1]{[#1]}%
\providecommand \BibitemOpen [0]{}%
\providecommand \bibitemStop [0]{}%
\providecommand \bibitemNoStop [0]{.\EOS\space}%
\providecommand \EOS [0]{\spacefactor3000\relax}%
\providecommand \BibitemShut  [1]{\csname bibitem#1\endcsname}%
\let\auto@bib@innerbib\@empty
\bibitem [{\citenamefont {Zayats}\ \emph {et~al.}(2005)\citenamefont {Zayats},
  \citenamefont {Smolyaninov},\ and\ \citenamefont {Maradudin}}]{Zayats2005}%
  \BibitemOpen
  \bibfield  {author} {\bibinfo {author} {\bibfnamefont {A.~V.}\ \bibnamefont
  {Zayats}}, \bibinfo {author} {\bibfnamefont {I.~I.}\ \bibnamefont
  {Smolyaninov}}, \ and\ \bibinfo {author} {\bibfnamefont {A.~A.}\ \bibnamefont
  {Maradudin}},\ }\href {\doibase
  http://dx.doi.org/10.1016/j.physrep.2004.11.001} {\bibfield  {journal}
  {\bibinfo  {journal} {Phys. Rep.}\ }\textbf {\bibinfo {volume} {408}},\
  \bibinfo {pages} {131 } (\bibinfo {year} {2005})}\BibitemShut {NoStop}%
\bibitem [{\citenamefont {Blackie}\ \emph {et~al.}(2009)\citenamefont
  {Blackie}, \citenamefont {Le~Ru},\ and\ \citenamefont
  {Etchegoin}}]{Blackie2009}%
  \BibitemOpen
  \bibfield  {author} {\bibinfo {author} {\bibfnamefont {E.~J.}\ \bibnamefont
  {Blackie}}, \bibinfo {author} {\bibfnamefont {E.~C.}\ \bibnamefont {Le~Ru}},
  \ and\ \bibinfo {author} {\bibfnamefont {P.~G.}\ \bibnamefont {Etchegoin}},\
  }\href {\doibase 10.1021/ja905319w} {\bibfield  {journal} {\bibinfo
  {journal} {J. Am. Chem. Soc.}\ }\textbf {\bibinfo {volume} {131}},\ \bibinfo
  {pages} {14466} (\bibinfo {year} {2009})},\ \Eprint
  {http://arxiv.org/abs/http://dx.doi.org/10.1021/ja905319w}
  {http://dx.doi.org/10.1021/ja905319w} \BibitemShut {NoStop}%
\bibitem [{\citenamefont {Gruenke}\ \emph {et~al.}(2016)\citenamefont
  {Gruenke}, \citenamefont {Cardinal}, \citenamefont {McAnally}, \citenamefont
  {Frontiera}, \citenamefont {Schatz},\ and\ \citenamefont {{Van
  Duyne}}}]{Gruenke2016}%
  \BibitemOpen
  \bibfield  {author} {\bibinfo {author} {\bibfnamefont {N.~L.}\ \bibnamefont
  {Gruenke}}, \bibinfo {author} {\bibfnamefont {M.~F.}\ \bibnamefont
  {Cardinal}}, \bibinfo {author} {\bibfnamefont {M.~O.}\ \bibnamefont
  {McAnally}}, \bibinfo {author} {\bibfnamefont {R.~R.}\ \bibnamefont
  {Frontiera}}, \bibinfo {author} {\bibfnamefont {G.~C.}\ \bibnamefont
  {Schatz}}, \ and\ \bibinfo {author} {\bibfnamefont {R.~P.}\ \bibnamefont
  {{Van Duyne}}},\ }\href {\doibase 10.1039/C5CS00763A} {\bibfield  {journal}
  {\bibinfo  {journal} {Chem. Soc. Rev.}\ }\textbf {\bibinfo {volume} {45}},\
  \bibinfo {pages} {2263} (\bibinfo {year} {2016})}\BibitemShut {NoStop}%
\bibitem [{\citenamefont {Stoerzinger}\ \emph {et~al.}(2011)\citenamefont
  {Stoerzinger}, \citenamefont {Lin},\ and\ \citenamefont {Odom}}]{Links2011}%
  \BibitemOpen
  \bibfield  {author} {\bibinfo {author} {\bibfnamefont {K.~A.}\ \bibnamefont
  {Stoerzinger}}, \bibinfo {author} {\bibfnamefont {J.~Y.}\ \bibnamefont
  {Lin}}, \ and\ \bibinfo {author} {\bibfnamefont {T.~W.}\ \bibnamefont
  {Odom}},\ }\href {\doibase 10.1039/c1sc00125f} {\bibfield  {journal}
  {\bibinfo  {journal} {Chem. Sci.}\ }\textbf {\bibinfo {volume} {2}},\
  \bibinfo {pages} {1435} (\bibinfo {year} {2011})}\BibitemShut {NoStop}%
\bibitem [{\citenamefont {{Le Ru}}\ \emph {et~al.}(2007)\citenamefont {{Le
  Ru}}, \citenamefont {Blackie}, \citenamefont {Meyer},\ and\ \citenamefont
  {Etchegoin}}]{Ru2007}%
  \BibitemOpen
  \bibfield  {author} {\bibinfo {author} {\bibfnamefont {E.~C.}\ \bibnamefont
  {{Le Ru}}}, \bibinfo {author} {\bibfnamefont {E.}~\bibnamefont {Blackie}},
  \bibinfo {author} {\bibfnamefont {M.}~\bibnamefont {Meyer}}, \ and\ \bibinfo
  {author} {\bibfnamefont {P.~G.}\ \bibnamefont {Etchegoin}},\ }\href {\doibase
  10.1021/jp0687908} {\bibfield  {journal} {\bibinfo  {journal} {J. Phys. Chem.
  C}\ }\textbf {\bibinfo {volume} {111}},\ \bibinfo {pages} {13794} (\bibinfo
  {year} {2007})}\BibitemShut {NoStop}%
\bibitem [{\citenamefont {Beams}\ \emph {et~al.}(2015)\citenamefont {Beams},
  \citenamefont {Can{\c{c}}ado}, \citenamefont {Jorio}, \citenamefont
  {Vamivakas},\ and\ \citenamefont {Novotny}}]{Beams2015}%
  \BibitemOpen
  \bibfield  {author} {\bibinfo {author} {\bibfnamefont {R.}~\bibnamefont
  {Beams}}, \bibinfo {author} {\bibfnamefont {L.~G.}\ \bibnamefont
  {Can{\c{c}}ado}}, \bibinfo {author} {\bibfnamefont {A.}~\bibnamefont
  {Jorio}}, \bibinfo {author} {\bibfnamefont {A.~N.}\ \bibnamefont
  {Vamivakas}}, \ and\ \bibinfo {author} {\bibfnamefont {L.}~\bibnamefont
  {Novotny}},\ }\href {\doibase 10.1088/0957-4484/26/17/175702} {\bibfield
  {journal} {\bibinfo  {journal} {Nanotechnology}\ }\textbf {\bibinfo {volume}
  {26}},\ \bibinfo {pages} {175702} (\bibinfo {year} {2015})}\BibitemShut
  {NoStop}%
\bibitem [{\citenamefont {Chen}\ \emph {et~al.}(2014)\citenamefont {Chen},
  \citenamefont {Hayazawa},\ and\ \citenamefont {Kawata}}]{Kawata2014}%
  \BibitemOpen
  \bibfield  {author} {\bibinfo {author} {\bibfnamefont {C.}~\bibnamefont
  {Chen}}, \bibinfo {author} {\bibfnamefont {N.}~\bibnamefont {Hayazawa}}, \
  and\ \bibinfo {author} {\bibfnamefont {S.}~\bibnamefont {Kawata}},\ }\href
  {\doibase 10.1038/ncomms4312} {\bibfield  {journal} {\bibinfo  {journal}
  {Nat. Commun.}\ }\textbf {\bibinfo {volume} {5}},\ \bibinfo {pages} {1}
  (\bibinfo {year} {2014})}\BibitemShut {NoStop}%
\bibitem [{\citenamefont {Sonntag}\ \emph {et~al.}(2014)\citenamefont
  {Sonntag}, \citenamefont {Pozzi}, \citenamefont {Jiang}, \citenamefont
  {Hersam},\ and\ \citenamefont {Duyne}}]{Sonntag2014}%
  \BibitemOpen
  \bibfield  {author} {\bibinfo {author} {\bibfnamefont {M.~D.}\ \bibnamefont
  {Sonntag}}, \bibinfo {author} {\bibfnamefont {E.~A.}\ \bibnamefont {Pozzi}},
  \bibinfo {author} {\bibfnamefont {N.}~\bibnamefont {Jiang}}, \bibinfo
  {author} {\bibfnamefont {M.~C.}\ \bibnamefont {Hersam}}, \ and\ \bibinfo
  {author} {\bibfnamefont {R.~P.~V.}\ \bibnamefont {Duyne}},\ }\href {\doibase
  10.1021/jz5015746} {\bibfield  {journal} {\bibinfo  {journal} {J. Phys. Chem.
  Lett.}\ }\textbf {\bibinfo {volume} {5}},\ \bibinfo {pages} {3125} (\bibinfo
  {year} {2014})}\BibitemShut {NoStop}%
\bibitem [{\citenamefont {Pozzi}\ \emph {et~al.}(2015)\citenamefont {Pozzi},
  \citenamefont {Zrimsek}, \citenamefont {Lethiec}, \citenamefont {Schatz},
  \citenamefont {Hersam},\ and\ \citenamefont {Duyne}}]{Pozzi2015}%
  \BibitemOpen
  \bibfield  {author} {\bibinfo {author} {\bibfnamefont {E.~A.}\ \bibnamefont
  {Pozzi}}, \bibinfo {author} {\bibfnamefont {A.~B.}\ \bibnamefont {Zrimsek}},
  \bibinfo {author} {\bibfnamefont {C.~M.}\ \bibnamefont {Lethiec}}, \bibinfo
  {author} {\bibfnamefont {G.~C.}\ \bibnamefont {Schatz}}, \bibinfo {author}
  {\bibfnamefont {M.~C.}\ \bibnamefont {Hersam}}, \ and\ \bibinfo {author}
  {\bibfnamefont {R.~P.~V.}\ \bibnamefont {Duyne}},\ }\href {\doibase
  10.1021/acs.jpcc.5b08054} {\bibfield  {journal} {\bibinfo  {journal} {J.
  Phys. Chem. C}\ }\textbf {\bibinfo {volume} {119}},\ \bibinfo {pages} {21116}
  (\bibinfo {year} {2015})},\ \Eprint
  {http://arxiv.org/abs/http://dx.doi.org/10.1021/acs.jpcc.5b08054}
  {http://dx.doi.org/10.1021/acs.jpcc.5b08054} \BibitemShut {NoStop}%
\bibitem [{\citenamefont {Nie}(1997)}]{Press2012}%
  \BibitemOpen
  \bibfield  {author} {\bibinfo {author} {\bibfnamefont {S.}~\bibnamefont
  {Nie}},\ }\href {\doibase 10.1126/science.275.5303.1102} {\bibfield
  {journal} {\bibinfo  {journal} {Science}\ }\textbf {\bibinfo {volume}
  {275}},\ \bibinfo {pages} {1102} (\bibinfo {year} {1997})}\BibitemShut
  {NoStop}%
\bibitem [{\citenamefont {Wu}\ and\ \citenamefont {Reinhard}(2014)}]{Wu2014}%
  \BibitemOpen
  \bibfield  {author} {\bibinfo {author} {\bibfnamefont {L.}~\bibnamefont
  {Wu}}\ and\ \bibinfo {author} {\bibfnamefont {B.~M.}\ \bibnamefont
  {Reinhard}},\ }\href {\doibase 10.1039/c3cs60340g} {\bibfield  {journal}
  {\bibinfo  {journal} {Chem. Soc. Rev.}\ }\textbf {\bibinfo {volume} {43}},\
  \bibinfo {pages} {3884} (\bibinfo {year} {2014})}\BibitemShut {NoStop}%
\bibitem [{\citenamefont {Cang}\ \emph {et~al.}(2015)\citenamefont {Cang},
  \citenamefont {Salandrino}, \citenamefont {Wang},\ and\ \citenamefont
  {Zhang}}]{Cang2015}%
  \BibitemOpen
  \bibfield  {author} {\bibinfo {author} {\bibfnamefont {H.}~\bibnamefont
  {Cang}}, \bibinfo {author} {\bibfnamefont {A.}~\bibnamefont {Salandrino}},
  \bibinfo {author} {\bibfnamefont {Y.}~\bibnamefont {Wang}}, \ and\ \bibinfo
  {author} {\bibfnamefont {X.}~\bibnamefont {Zhang}},\ }\href {\doibase
  10.1038/ncomms8942} {\bibfield  {journal} {\bibinfo  {journal} {Nat.
  Commun.}\ }\textbf {\bibinfo {volume} {6}},\ \bibinfo {pages} {1} (\bibinfo
  {year} {2015})}\BibitemShut {NoStop}%
\bibitem [{\citenamefont {Smolyaninov}(2008)}]{Smolyaninov2015}%
  \BibitemOpen
  \bibfield  {author} {\bibinfo {author} {\bibfnamefont {I.~I.}\ \bibnamefont
  {Smolyaninov}},\ }\href {\doibase 10.2976/1.2912559} {\bibfield  {journal}
  {\bibinfo  {journal} {HFSP J}\ }\textbf {\bibinfo {volume} {2}},\ \bibinfo
  {pages} {129} (\bibinfo {year} {2008})}\BibitemShut {NoStop}%
\bibitem [{\citenamefont {Esteves-L{\'{o}}pez}\ \emph
  {et~al.}(2015)\citenamefont {Esteves-L{\'{o}}pez}, \citenamefont
  {Pastawski},\ and\ \citenamefont {Bustos-Mar{\'{u}}n}}]{Bustos-Marun2015}%
  \BibitemOpen
  \bibfield  {author} {\bibinfo {author} {\bibfnamefont {N.}~\bibnamefont
  {Esteves-L{\'{o}}pez}}, \bibinfo {author} {\bibfnamefont {H.~M.}\
  \bibnamefont {Pastawski}}, \ and\ \bibinfo {author} {\bibfnamefont {R.~A.}\
  \bibnamefont {Bustos-Mar{\'{u}}n}},\ }\href {\doibase
  10.1088/0953-8984/27/12/125301} {\bibfield  {journal} {\bibinfo  {journal}
  {J. Phys.: Condens. Matter}\ }\textbf {\bibinfo {volume} {27}},\ \bibinfo
  {pages} {125301} (\bibinfo {year} {2015})},\ \Eprint
  {http://arxiv.org/abs/arXiv:1502.01196v1} {arXiv:arXiv:1502.01196v1}
  \BibitemShut {NoStop}%
\bibitem [{\citenamefont {Eisenthal}(2006)}]{Eisenthal2006}%
  \BibitemOpen
  \bibfield  {author} {\bibinfo {author} {\bibfnamefont {K.~B.}\ \bibnamefont
  {Eisenthal}},\ }\href {\doibase 10.1021/cr0403685} {\bibfield  {journal}
  {\bibinfo  {journal} {Chem. Rev.}\ }\textbf {\bibinfo {volume} {106}},\
  \bibinfo {pages} {1462} (\bibinfo {year} {2006})}\BibitemShut {NoStop}%
\bibitem [{\citenamefont {Fan}\ \emph {et~al.}(2006)\citenamefont {Fan},
  \citenamefont {Zhang}, \citenamefont {Panoiu}, \citenamefont {Abdenour},
  \citenamefont {Krishna}, \citenamefont {Osgood}, \citenamefont {Malloy},\
  and\ \citenamefont {Brueck}}]{Fan2006}%
  \BibitemOpen
  \bibfield  {author} {\bibinfo {author} {\bibfnamefont {W.}~\bibnamefont
  {Fan}}, \bibinfo {author} {\bibfnamefont {S.}~\bibnamefont {Zhang}}, \bibinfo
  {author} {\bibfnamefont {N.~C.}\ \bibnamefont {Panoiu}}, \bibinfo {author}
  {\bibfnamefont {A.}~\bibnamefont {Abdenour}}, \bibinfo {author}
  {\bibfnamefont {S.}~\bibnamefont {Krishna}}, \bibinfo {author} {\bibfnamefont
  {R.~M.}\ \bibnamefont {Osgood}}, \bibinfo {author} {\bibfnamefont {K.~J.}\
  \bibnamefont {Malloy}}, \ and\ \bibinfo {author} {\bibfnamefont {S.~R.~J.}\
  \bibnamefont {Brueck}},\ }\href {\doibase 10.1021/nl0604457} {\bibfield
  {journal} {\bibinfo  {journal} {Nano Lett.}\ }\textbf {\bibinfo {volume}
  {6}},\ \bibinfo {pages} {1027} (\bibinfo {year} {2006})}\BibitemShut
  {NoStop}%
\bibitem [{\citenamefont {Kim}\ \emph {et~al.}(2008)\citenamefont {Kim},
  \citenamefont {Jin}, \citenamefont {Kim}, \citenamefont {Park}, \citenamefont
  {Kim},\ and\ \citenamefont {Kim}}]{Kim2008}%
  \BibitemOpen
  \bibfield  {author} {\bibinfo {author} {\bibfnamefont {S.}~\bibnamefont
  {Kim}}, \bibinfo {author} {\bibfnamefont {J.}~\bibnamefont {Jin}}, \bibinfo
  {author} {\bibfnamefont {Y.-J.}\ \bibnamefont {Kim}}, \bibinfo {author}
  {\bibfnamefont {I.-Y.}\ \bibnamefont {Park}}, \bibinfo {author}
  {\bibfnamefont {Y.}~\bibnamefont {Kim}}, \ and\ \bibinfo {author}
  {\bibfnamefont {S.-W.}\ \bibnamefont {Kim}},\ }\href {\doibase
  10.1038/nature07012} {\bibfield  {journal} {\bibinfo  {journal} {Nature}\
  }\textbf {\bibinfo {volume} {453}},\ \bibinfo {pages} {757} (\bibinfo {year}
  {2008})}\BibitemShut {NoStop}%
\bibitem [{\citenamefont {Kim}\ and\ \citenamefont {Rho}(2015)}]{Kim2015}%
  \BibitemOpen
  \bibfield  {author} {\bibinfo {author} {\bibfnamefont {M.}~\bibnamefont
  {Kim}}\ and\ \bibinfo {author} {\bibfnamefont {J.}~\bibnamefont {Rho}},\
  }\href {\doibase 10.1186/s40580-015-0053-7} {\bibfield  {journal} {\bibinfo
  {journal} {Nano Convergence}\ }\textbf {\bibinfo {volume} {2}},\ \bibinfo
  {pages} {22} (\bibinfo {year} {2015})}\BibitemShut {NoStop}%
\bibitem [{\citenamefont {Boltasseva}\ and\ \citenamefont
  {Atwater}(2011)}]{AlexandraBoltasseva2012}%
  \BibitemOpen
  \bibfield  {author} {\bibinfo {author} {\bibfnamefont {A.}~\bibnamefont
  {Boltasseva}}\ and\ \bibinfo {author} {\bibfnamefont {H.~a.}\ \bibnamefont
  {Atwater}},\ }\href {\doibase 10.1126/science.1198258} {\bibfield  {journal}
  {\bibinfo  {journal} {Science}\ }\textbf {\bibinfo {volume} {331}},\ \bibinfo
  {pages} {290} (\bibinfo {year} {2011})}\BibitemShut {NoStop}%
\bibitem [{\citenamefont {Arnold}\ \emph {et~al.}(2013)\citenamefont {Arnold},
  \citenamefont {Ding}, \citenamefont {Hrelescu},\ and\ \citenamefont
  {Klar}}]{Arnold2013}%
  \BibitemOpen
  \bibfield  {author} {\bibinfo {author} {\bibfnamefont {N.}~\bibnamefont
  {Arnold}}, \bibinfo {author} {\bibfnamefont {B.}~\bibnamefont {Ding}},
  \bibinfo {author} {\bibfnamefont {C.}~\bibnamefont {Hrelescu}}, \ and\
  \bibinfo {author} {\bibfnamefont {T.~a.}\ \bibnamefont {Klar}},\ }\href
  {\doibase 10.3762/bjnano.4.110} {\bibfield  {journal} {\bibinfo  {journal}
  {Beilstein J. Nanotechnol.}\ }\textbf {\bibinfo {volume} {4}},\ \bibinfo
  {pages} {974} (\bibinfo {year} {2013})}\BibitemShut {NoStop}%
\bibitem [{Gor(2007)}]{Gordon2007}%
  \BibitemOpen
  \href {\doibase 10.1364/OE.15.002622} {\bibfield  {journal} {\bibinfo
  {journal} {Opt. Express}\ }\textbf {\bibinfo {volume} {15}},\ \bibinfo
  {pages} {2622} (\bibinfo {year} {2007})},\ \Eprint
  {http://arxiv.org/abs/0612192} {0612192 [physics]} \BibitemShut {NoStop}%
\bibitem [{\citenamefont {Bergman}\ and\ \citenamefont
  {Stockman}(2003)}]{Bergman2003}%
  \BibitemOpen
  \bibfield  {author} {\bibinfo {author} {\bibfnamefont {D.~J.}\ \bibnamefont
  {Bergman}}\ and\ \bibinfo {author} {\bibfnamefont {M.~I.}\ \bibnamefont
  {Stockman}},\ }\href {\doibase 10.1103/PhysRevLett.90.027402} {\bibfield
  {journal} {\bibinfo  {journal} {Phys. Rev. Lett.}\ }\textbf {\bibinfo
  {volume} {90}},\ \bibinfo {pages} {027402} (\bibinfo {year}
  {2003})}\BibitemShut {NoStop}%
\bibitem [{\citenamefont {Liu}\ \emph {et~al.}(2011)\citenamefont {Liu},
  \citenamefont {Li}, \citenamefont {Zhou}, \citenamefont {Gan},\ and\
  \citenamefont {Li}}]{Liu2011}%
  \BibitemOpen
  \bibfield  {author} {\bibinfo {author} {\bibfnamefont {S.-Y.}\ \bibnamefont
  {Liu}}, \bibinfo {author} {\bibfnamefont {J.}~\bibnamefont {Li}}, \bibinfo
  {author} {\bibfnamefont {F.}~\bibnamefont {Zhou}}, \bibinfo {author}
  {\bibfnamefont {L.}~\bibnamefont {Gan}}, \ and\ \bibinfo {author}
  {\bibfnamefont {Z.-Y.}\ \bibnamefont {Li}},\ }\href {\doibase
  10.1364/OL.36.001296} {\bibfield  {journal} {\bibinfo  {journal} {Opt.
  Lett.}\ }\textbf {\bibinfo {volume} {36}},\ \bibinfo {pages} {1296} (\bibinfo
  {year} {2011})}\BibitemShut {NoStop}%
\bibitem [{\citenamefont {Yu}\ \emph {et~al.}(2015)\citenamefont {Yu},
  \citenamefont {Jiang},\ and\ \citenamefont {Wu}}]{Yu2015}%
  \BibitemOpen
  \bibfield  {author} {\bibinfo {author} {\bibfnamefont {H.}~\bibnamefont
  {Yu}}, \bibinfo {author} {\bibfnamefont {S.}~\bibnamefont {Jiang}}, \ and\
  \bibinfo {author} {\bibfnamefont {D.}~\bibnamefont {Wu}},\ }\href {\doibase
  10.1063/1.4918310} {\bibfield  {journal} {\bibinfo  {journal} {J. Appl.
  Phys.}\ }\textbf {\bibinfo {volume} {117}},\ \bibinfo {pages} {153101}
  (\bibinfo {year} {2015})}\BibitemShut {NoStop}%
\bibitem [{\citenamefont {Rasskazov}\ \emph {et~al.}(2013)\citenamefont
  {Rasskazov}, \citenamefont {Karpov},\ and\ \citenamefont
  {Markel}}]{Rasskazov2013}%
  \BibitemOpen
  \bibfield  {author} {\bibinfo {author} {\bibfnamefont {I.~L.}\ \bibnamefont
  {Rasskazov}}, \bibinfo {author} {\bibfnamefont {S.~V.}\ \bibnamefont
  {Karpov}}, \ and\ \bibinfo {author} {\bibfnamefont {V.~a.}\ \bibnamefont
  {Markel}},\ }\href {\doibase 10.1364/OL.38.004743} {\bibfield  {journal}
  {\bibinfo  {journal} {Opt. Lett.}\ }\textbf {\bibinfo {volume} {38}},\
  \bibinfo {pages} {4743} (\bibinfo {year} {2013})}\BibitemShut {NoStop}%
\bibitem [{\citenamefont {Citrin}(2006)}]{Citrin2006}%
  \BibitemOpen
  \bibfield  {author} {\bibinfo {author} {\bibfnamefont {D.~S.}\ \bibnamefont
  {Citrin}},\ }\href {\doibase 10.1364/OL.31.000098} {\bibfield  {journal}
  {\bibinfo  {journal} {Opt. Lett.}\ }\textbf {\bibinfo {volume} {31}},\
  \bibinfo {pages} {98} (\bibinfo {year} {2006})}\BibitemShut {NoStop}%
\bibitem [{\citenamefont {Nugroho}\ \emph {et~al.}(2015)\citenamefont
  {Nugroho}, \citenamefont {Malyshev},\ and\ \citenamefont
  {Knoester}}]{Nugroho2015}%
  \BibitemOpen
  \bibfield  {author} {\bibinfo {author} {\bibfnamefont {B.~S.}\ \bibnamefont
  {Nugroho}}, \bibinfo {author} {\bibfnamefont {V.~A.}\ \bibnamefont
  {Malyshev}}, \ and\ \bibinfo {author} {\bibfnamefont {J.}~\bibnamefont
  {Knoester}},\ }\href {\doibase 10.1103/PhysRevB.92.165432} {\bibfield
  {journal} {\bibinfo  {journal} {Phys. Rev. B}\ }\textbf {\bibinfo {volume}
  {92}},\ \bibinfo {pages} {1} (\bibinfo {year} {2015})},\ \Eprint
  {http://arxiv.org/abs/1508.04197} {arXiv:1508.04197} \BibitemShut {NoStop}%
\bibitem [{\citenamefont {Fang}\ \emph {et~al.}(2009)\citenamefont {Fang},
  \citenamefont {Koschny}, \citenamefont {Wegener},\ and\ \citenamefont
  {Soukoulis}}]{Fang2009}%
  \BibitemOpen
  \bibfield  {author} {\bibinfo {author} {\bibfnamefont {a.}~\bibnamefont
  {Fang}}, \bibinfo {author} {\bibfnamefont {T.}~\bibnamefont {Koschny}},
  \bibinfo {author} {\bibfnamefont {M.}~\bibnamefont {Wegener}}, \ and\
  \bibinfo {author} {\bibfnamefont {C.~M.}\ \bibnamefont {Soukoulis}},\ }\href
  {\doibase 10.1103/PhysRevB.79.241104} {\bibfield  {journal} {\bibinfo
  {journal} {Phys. Rev. B: Condens. Matter}\ }\textbf {\bibinfo {volume}
  {79}},\ \bibinfo {pages} {8} (\bibinfo {year} {2009})},\ \Eprint
  {http://arxiv.org/abs/0907.0888} {arXiv:0907.0888} \BibitemShut {NoStop}%
\bibitem [{\citenamefont {Hamm}\ \emph {et~al.}(2011)\citenamefont {Hamm},
  \citenamefont {Wuestner}, \citenamefont {Tsakmakidis},\ and\ \citenamefont
  {Hess}}]{Hamm2011}%
  \BibitemOpen
  \bibfield  {author} {\bibinfo {author} {\bibfnamefont {J.~M.}\ \bibnamefont
  {Hamm}}, \bibinfo {author} {\bibfnamefont {S.}~\bibnamefont {Wuestner}},
  \bibinfo {author} {\bibfnamefont {K.~L.}\ \bibnamefont {Tsakmakidis}}, \ and\
  \bibinfo {author} {\bibfnamefont {O.}~\bibnamefont {Hess}},\ }\href {\doibase
  10.1103/PhysRevLett.107.167405} {\bibfield  {journal} {\bibinfo  {journal}
  {Phys. Rev. Lett.}\ }\textbf {\bibinfo {volume} {107}},\ \bibinfo {pages} {1}
  (\bibinfo {year} {2011})},\ \Eprint {http://arxiv.org/abs/arXiv:1109.4411v1}
  {arXiv:arXiv:1109.4411v1} \BibitemShut {NoStop}%
\bibitem [{\citenamefont {Wuestner}\ \emph {et~al.}(2011)\citenamefont
  {Wuestner}, \citenamefont {Pusch}, \citenamefont {Tsakmakidis}, \citenamefont
  {Hamm},\ and\ \citenamefont {Hess}}]{Wuestner2011}%
  \BibitemOpen
  \bibfield  {author} {\bibinfo {author} {\bibfnamefont {S.}~\bibnamefont
  {Wuestner}}, \bibinfo {author} {\bibfnamefont {A.}~\bibnamefont {Pusch}},
  \bibinfo {author} {\bibfnamefont {K.~L.}\ \bibnamefont {Tsakmakidis}},
  \bibinfo {author} {\bibfnamefont {J.~M.}\ \bibnamefont {Hamm}}, \ and\
  \bibinfo {author} {\bibfnamefont {O.}~\bibnamefont {Hess}},\ }\href {\doibase
  10.1098/rsta.2011.0140} {\bibfield  {journal} {\bibinfo  {journal} {Phil.
  Trans. R. Soc. A}\ }\textbf {\bibinfo {volume} {369}},\ \bibinfo {pages}
  {3525} (\bibinfo {year} {2011})},\ \Eprint {http://arxiv.org/abs/1012.1576}
  {arXiv:1012.1576} \BibitemShut {NoStop}%
\bibitem [{\citenamefont {Wuestner}\ \emph {et~al.}(2010)\citenamefont
  {Wuestner}, \citenamefont {Pusch}, \citenamefont {Tsakmakidis}, \citenamefont
  {Hamm},\ and\ \citenamefont {Hess}}]{Wuestner2010}%
  \BibitemOpen
  \bibfield  {author} {\bibinfo {author} {\bibfnamefont {S.}~\bibnamefont
  {Wuestner}}, \bibinfo {author} {\bibfnamefont {A.}~\bibnamefont {Pusch}},
  \bibinfo {author} {\bibfnamefont {K.~L.}\ \bibnamefont {Tsakmakidis}},
  \bibinfo {author} {\bibfnamefont {J.~M.}\ \bibnamefont {Hamm}}, \ and\
  \bibinfo {author} {\bibfnamefont {O.}~\bibnamefont {Hess}},\ }\href {\doibase
  10.1103/PhysRevLett.105.127401} {\bibfield  {journal} {\bibinfo  {journal}
  {Phys. Rev. Lett.}\ }\textbf {\bibinfo {volume} {105}},\ \bibinfo {pages} {1}
  (\bibinfo {year} {2010})},\ \Eprint {http://arxiv.org/abs/1006.5926}
  {arXiv:1006.5926} \BibitemShut {NoStop}%
\bibitem [{\citenamefont {{De Luca}}\ \emph {et~al.}(2014)\citenamefont {{De
  Luca}}, \citenamefont {Dhama}, \citenamefont {Rashed}, \citenamefont
  {Coutant}, \citenamefont {Ravaine}, \citenamefont {Barois}, \citenamefont
  {Infusino},\ and\ \citenamefont {Strangi}}]{DeLuca2014}%
  \BibitemOpen
  \bibfield  {author} {\bibinfo {author} {\bibfnamefont {a.}~\bibnamefont {{De
  Luca}}}, \bibinfo {author} {\bibfnamefont {R.}~\bibnamefont {Dhama}},
  \bibinfo {author} {\bibfnamefont {a.~R.}\ \bibnamefont {Rashed}}, \bibinfo
  {author} {\bibfnamefont {C.}~\bibnamefont {Coutant}}, \bibinfo {author}
  {\bibfnamefont {S.}~\bibnamefont {Ravaine}}, \bibinfo {author} {\bibfnamefont
  {P.}~\bibnamefont {Barois}}, \bibinfo {author} {\bibfnamefont
  {M.}~\bibnamefont {Infusino}}, \ and\ \bibinfo {author} {\bibfnamefont
  {G.}~\bibnamefont {Strangi}},\ }\href {\doibase 10.1063/1.4868105} {\bibfield
   {journal} {\bibinfo  {journal} {Appl. Phys. Lett}\ }\textbf {\bibinfo
  {volume} {104}} (\bibinfo {year} {2014}),\ 10.1063/1.4868105}\BibitemShut
  {NoStop}%
\bibitem [{\citenamefont {{De Luca}}\ \emph {et~al.}(2012)\citenamefont {{De
  Luca}}, \citenamefont {Ferrie}, \citenamefont {Ravaine}, \citenamefont {{La
  Deda}}, \citenamefont {Infusino}, \citenamefont {Rashed}, \citenamefont
  {Veltri}, \citenamefont {Aradian}, \citenamefont {Scaramuzza},\ and\
  \citenamefont {Strangi}}]{DeLuca2012}%
  \BibitemOpen
  \bibfield  {author} {\bibinfo {author} {\bibfnamefont {A.}~\bibnamefont {{De
  Luca}}}, \bibinfo {author} {\bibfnamefont {M.}~\bibnamefont {Ferrie}},
  \bibinfo {author} {\bibfnamefont {S.}~\bibnamefont {Ravaine}}, \bibinfo
  {author} {\bibfnamefont {M.}~\bibnamefont {{La Deda}}}, \bibinfo {author}
  {\bibfnamefont {M.}~\bibnamefont {Infusino}}, \bibinfo {author}
  {\bibfnamefont {A.~R.}\ \bibnamefont {Rashed}}, \bibinfo {author}
  {\bibfnamefont {A.}~\bibnamefont {Veltri}}, \bibinfo {author} {\bibfnamefont
  {A.}~\bibnamefont {Aradian}}, \bibinfo {author} {\bibfnamefont
  {N.}~\bibnamefont {Scaramuzza}}, \ and\ \bibinfo {author} {\bibfnamefont
  {G.}~\bibnamefont {Strangi}},\ }\href {\doibase 10.1039/c2jm30341h}
  {\bibfield  {journal} {\bibinfo  {journal} {J. Mater. Chem.}\ }\textbf
  {\bibinfo {volume} {22}},\ \bibinfo {pages} {8846} (\bibinfo {year}
  {2012})}\BibitemShut {NoStop}%
\bibitem [{\citenamefont {Noginov}\ \emph {et~al.}(2009)\citenamefont
  {Noginov}, \citenamefont {Zhu}, \citenamefont {Belgrave}, \citenamefont
  {Bakker}, \citenamefont {Shalaev}, \citenamefont {Narimanov}, \citenamefont
  {Stout}, \citenamefont {Herz}, \citenamefont {Suteewong},\ and\ \citenamefont
  {Wiesner}}]{Noginov2009}%
  \BibitemOpen
  \bibfield  {author} {\bibinfo {author} {\bibfnamefont {M.~A.}\ \bibnamefont
  {Noginov}}, \bibinfo {author} {\bibfnamefont {G.}~\bibnamefont {Zhu}},
  \bibinfo {author} {\bibfnamefont {A.~M.}\ \bibnamefont {Belgrave}}, \bibinfo
  {author} {\bibfnamefont {R.}~\bibnamefont {Bakker}}, \bibinfo {author}
  {\bibfnamefont {V.~M.}\ \bibnamefont {Shalaev}}, \bibinfo {author}
  {\bibfnamefont {E.~E.}\ \bibnamefont {Narimanov}}, \bibinfo {author}
  {\bibfnamefont {S.}~\bibnamefont {Stout}}, \bibinfo {author} {\bibfnamefont
  {E.}~\bibnamefont {Herz}}, \bibinfo {author} {\bibfnamefont {T.}~\bibnamefont
  {Suteewong}}, \ and\ \bibinfo {author} {\bibfnamefont {U.}~\bibnamefont
  {Wiesner}},\ }\href {\doibase 10.1038/nature08318} {\bibfield  {journal}
  {\bibinfo  {journal} {Nature}\ }\textbf {\bibinfo {volume} {460}},\ \bibinfo
  {pages} {1110} (\bibinfo {year} {2009})}\BibitemShut {NoStop}%
\bibitem [{\citenamefont {Infusino}\ \emph {et~al.}(2014)\citenamefont
  {Infusino}, \citenamefont {{De Luca}}, \citenamefont {Veltri}, \citenamefont
  {V{\'{a}}zquez-V{\'{a}}zquez}, \citenamefont {Correa-Duarte}, \citenamefont
  {Dhama},\ and\ \citenamefont {Strangi}}]{Infusino2014}%
  \BibitemOpen
  \bibfield  {author} {\bibinfo {author} {\bibfnamefont {M.}~\bibnamefont
  {Infusino}}, \bibinfo {author} {\bibfnamefont {A.}~\bibnamefont {{De Luca}}},
  \bibinfo {author} {\bibfnamefont {A.}~\bibnamefont {Veltri}}, \bibinfo
  {author} {\bibfnamefont {C.}~\bibnamefont {V{\'{a}}zquez-V{\'{a}}zquez}},
  \bibinfo {author} {\bibfnamefont {M.~A.}\ \bibnamefont {Correa-Duarte}},
  \bibinfo {author} {\bibfnamefont {R.}~\bibnamefont {Dhama}}, \ and\ \bibinfo
  {author} {\bibfnamefont {G.}~\bibnamefont {Strangi}},\ }\href {\doibase
  10.1021/ph400174p} {\bibfield  {journal} {\bibinfo  {journal} {ACS
  Photonics}\ }\textbf {\bibinfo {volume} {1}},\ \bibinfo {pages} {371}
  (\bibinfo {year} {2014})}\BibitemShut {NoStop}%
\bibitem [{\citenamefont {Noginov}\ \emph {et~al.}(2007)\citenamefont
  {Noginov}, \citenamefont {Zhu}, \citenamefont {Bahoura}, \citenamefont
  {Adegoke}, \citenamefont {Small}, \citenamefont {Ritzo}, \citenamefont
  {Drachev},\ and\ \citenamefont {Shalaev}}]{Noginov2007}%
  \BibitemOpen
  \bibfield  {author} {\bibinfo {author} {\bibfnamefont {M.~a.}\ \bibnamefont
  {Noginov}}, \bibinfo {author} {\bibfnamefont {G.}~\bibnamefont {Zhu}},
  \bibinfo {author} {\bibfnamefont {M.}~\bibnamefont {Bahoura}}, \bibinfo
  {author} {\bibfnamefont {J.}~\bibnamefont {Adegoke}}, \bibinfo {author}
  {\bibfnamefont {C.}~\bibnamefont {Small}}, \bibinfo {author} {\bibfnamefont
  {B.~a.}\ \bibnamefont {Ritzo}}, \bibinfo {author} {\bibfnamefont {V.~P.}\
  \bibnamefont {Drachev}}, \ and\ \bibinfo {author} {\bibfnamefont {V.~M.}\
  \bibnamefont {Shalaev}},\ }\href {\doibase 10.1007/s00340-006-2401-0}
  {\bibfield  {journal} {\bibinfo  {journal} {Appl. Phys. B: Lasers Opt.}\
  }\textbf {\bibinfo {volume} {86}},\ \bibinfo {pages} {455} (\bibinfo {year}
  {2007})}\BibitemShut {NoStop}%
\bibitem [{\citenamefont {Zhou}\ \emph {et~al.}(2013)\citenamefont {Zhou},
  \citenamefont {Dridi}, \citenamefont {Suh}, \citenamefont {Kim},
  \citenamefont {Co}, \citenamefont {Wasielewski}, \citenamefont {Schatz},\
  and\ \citenamefont {Odom}}]{Zhou2013}%
  \BibitemOpen
  \bibfield  {author} {\bibinfo {author} {\bibfnamefont {W.}~\bibnamefont
  {Zhou}}, \bibinfo {author} {\bibfnamefont {M.}~\bibnamefont {Dridi}},
  \bibinfo {author} {\bibfnamefont {J.~Y.}\ \bibnamefont {Suh}}, \bibinfo
  {author} {\bibfnamefont {C.~H.}\ \bibnamefont {Kim}}, \bibinfo {author}
  {\bibfnamefont {D.~T.}\ \bibnamefont {Co}}, \bibinfo {author} {\bibfnamefont
  {M.~R.}\ \bibnamefont {Wasielewski}}, \bibinfo {author} {\bibfnamefont
  {G.~C.}\ \bibnamefont {Schatz}}, \ and\ \bibinfo {author} {\bibfnamefont
  {T.~W.}\ \bibnamefont {Odom}},\ }\href {\doibase 10.1038/nnano.2013.99}
  {\bibfield  {journal} {\bibinfo  {journal} {Nat. Nanotechnol.}\ }\textbf
  {\bibinfo {volume} {8}},\ \bibinfo {pages} {506} (\bibinfo {year} {2013})},\
  \Eprint {http://arxiv.org/abs/NNANO.2013.99} {arXiv:NNANO.2013.99 [10.1038]}
  \BibitemShut {NoStop}%
\bibitem [{\citenamefont {Stockman}(2009)}]{Stockman2009}%
  \BibitemOpen
  \bibfield  {author} {\bibinfo {author} {\bibfnamefont {M.~I.}\ \bibnamefont
  {Stockman}},\ }\href@noop {} {\bibfield  {journal} {\bibinfo  {journal} {J.
  Opt.}\ ,\ \bibinfo {pages} {1}} (\bibinfo {year} {2009})}\BibitemShut
  {NoStop}%
\bibitem [{\citenamefont {Bustos-Mar{\'{u}}n}\ \emph
  {et~al.}(2014)\citenamefont {Bustos-Mar{\'{u}}n}, \citenamefont {Dente},
  \citenamefont {Coronado},\ and\ \citenamefont
  {Pastawski}}]{Bustos-Marun2014}%
  \BibitemOpen
  \bibfield  {author} {\bibinfo {author} {\bibfnamefont {R.~A.}\ \bibnamefont
  {Bustos-Mar{\'{u}}n}}, \bibinfo {author} {\bibfnamefont {A.~D.}\ \bibnamefont
  {Dente}}, \bibinfo {author} {\bibfnamefont {E.~A.}\ \bibnamefont {Coronado}},
  \ and\ \bibinfo {author} {\bibfnamefont {H.~M.}\ \bibnamefont {Pastawski}},\
  }\href@noop {} {\bibfield  {journal} {\bibinfo  {journal} {Plasmonics}\
  }\textbf {\bibinfo {volume} {9}},\ \bibinfo {pages} {925} (\bibinfo {year}
  {2014})}\BibitemShut {NoStop}%
\bibitem [{\citenamefont {Lawandy}(2004)}]{Lawandy2004}%
  \BibitemOpen
  \bibfield  {author} {\bibinfo {author} {\bibfnamefont {N.~M.}\ \bibnamefont
  {Lawandy}},\ }\href {\doibase 10.1063/1.1825058} {\bibfield  {journal}
  {\bibinfo  {journal} {Appl. Phys. Lett.}\ }\textbf {\bibinfo {volume} {85}},\
  \bibinfo {pages} {5040} (\bibinfo {year} {2004})}\BibitemShut {NoStop}%
\bibitem [{\citenamefont {Li}\ and\ \citenamefont {Xia}(2010)}]{Li2010}%
  \BibitemOpen
  \bibfield  {author} {\bibinfo {author} {\bibfnamefont {Z.-Y.}\ \bibnamefont
  {Li}}\ and\ \bibinfo {author} {\bibfnamefont {Y.}~\bibnamefont {Xia}},\
  }\href {\doibase 10.1021/nl903409x} {\bibfield  {journal} {\bibinfo
  {journal} {Nano Lett.}\ }\textbf {\bibinfo {volume} {10}},\ \bibinfo {pages}
  {243} (\bibinfo {year} {2010})}\BibitemShut {NoStop}%
\bibitem [{\citenamefont {Arnold}\ \emph {et~al.}(2016)\citenamefont {Arnold},
  \citenamefont {Hrelescu},\ and\ \citenamefont {Klar}}]{Arnold2016}%
  \BibitemOpen
  \bibfield  {author} {\bibinfo {author} {\bibfnamefont {N.}~\bibnamefont
  {Arnold}}, \bibinfo {author} {\bibfnamefont {C.}~\bibnamefont {Hrelescu}}, \
  and\ \bibinfo {author} {\bibfnamefont {T.}~\bibnamefont {Klar}},\ }\href@noop
  {} {\bibfield  {journal} {\bibinfo  {journal} {Ann. Phys. (Berlin)}\ }\textbf
  {\bibinfo {volume} {528}},\ \bibinfo {pages} {295–306} (\bibinfo {year}
  {2016})}\BibitemShut {NoStop}%
\bibitem [{\citenamefont {Westcott}\ \emph {et~al.}(1998)\citenamefont
  {Westcott}, \citenamefont {Oldenburg}, \citenamefont {Lee},\ and\
  \citenamefont {Halas}}]{Westcott1998}%
  \BibitemOpen
  \bibfield  {author} {\bibinfo {author} {\bibfnamefont {S.~L.}\ \bibnamefont
  {Westcott}}, \bibinfo {author} {\bibfnamefont {S.~J.}\ \bibnamefont
  {Oldenburg}}, \bibinfo {author} {\bibfnamefont {T.~R.}\ \bibnamefont {Lee}},
  \ and\ \bibinfo {author} {\bibfnamefont {N.~J.}\ \bibnamefont {Halas}},\
  }\href {\doibase 10.1021/la980380q} {\bibfield  {journal} {\bibinfo
  {journal} {Langmuir}\ }\textbf {\bibinfo {volume} {14}},\ \bibinfo {pages}
  {5396} (\bibinfo {year} {1998})}\BibitemShut {NoStop}%
\bibitem [{\citenamefont {Phonthammachai}\ \emph {et~al.}(2008)\citenamefont
  {Phonthammachai}, \citenamefont {Kah}, \citenamefont {Jun}, \citenamefont
  {Sheppard}, \citenamefont {Olivo}, \citenamefont {Mhaisalkar},\ and\
  \citenamefont {White}}]{Sheppard2008}%
  \BibitemOpen
  \bibfield  {author} {\bibinfo {author} {\bibfnamefont {N.}~\bibnamefont
  {Phonthammachai}}, \bibinfo {author} {\bibfnamefont {J.~C.~Y.}\ \bibnamefont
  {Kah}}, \bibinfo {author} {\bibfnamefont {G.}~\bibnamefont {Jun}}, \bibinfo
  {author} {\bibfnamefont {C.~J.~R.}\ \bibnamefont {Sheppard}}, \bibinfo
  {author} {\bibfnamefont {M.~C.}\ \bibnamefont {Olivo}}, \bibinfo {author}
  {\bibfnamefont {S.~G.}\ \bibnamefont {Mhaisalkar}}, \ and\ \bibinfo {author}
  {\bibfnamefont {T.~J.}\ \bibnamefont {White}},\ }\href {\doibase
  10.1021/la703580r} {\bibfield  {journal} {\bibinfo  {journal} {Langmuir}\
  }\textbf {\bibinfo {volume} {24}},\ \bibinfo {pages} {5109} (\bibinfo {year}
  {2008})}\BibitemShut {NoStop}%
\bibitem [{\citenamefont {Bohren}\ and\ \citenamefont
  {Huffman}(1983)}]{craig1983absorption}%
  \BibitemOpen
  \bibfield  {author} {\bibinfo {author} {\bibfnamefont {C.~F.}\ \bibnamefont
  {Bohren}}\ and\ \bibinfo {author} {\bibfnamefont {D.}~\bibnamefont
  {Huffman}},\ }\href {https://books.google.com.ar/books?id=S1RCZ8BjgN0C}
  {\emph {\bibinfo {title} {Absorption and scattering of light by small
  particles}}},\ Wiley science paperback series\ (\bibinfo  {publisher}
  {Wiley},\ \bibinfo {year} {1983})\BibitemShut {NoStop}%
\bibitem [{\citenamefont {Maier}(2007)}]{maier2007plasmonics}%
  \BibitemOpen
  \bibfield  {author} {\bibinfo {author} {\bibfnamefont {S.}~\bibnamefont
  {Maier}},\ }\href {https://books.google.ru/books?id=yT2ux7TmDc8C} {\emph
  {\bibinfo {title} {Plasmonics: Fundamentals and Applications}}}\ (\bibinfo
  {publisher} {Springer US},\ \bibinfo {year} {2007})\BibitemShut {NoStop}%
\bibitem [{\citenamefont {Calander}\ \emph {et~al.}(2012)\citenamefont
  {Calander}, \citenamefont {Jin},\ and\ \citenamefont
  {Goldys}}]{Calander2012}%
  \BibitemOpen
  \bibfield  {author} {\bibinfo {author} {\bibfnamefont {N.}~\bibnamefont
  {Calander}}, \bibinfo {author} {\bibfnamefont {D.}~\bibnamefont {Jin}}, \
  and\ \bibinfo {author} {\bibfnamefont {E.~M.}\ \bibnamefont {Goldys}},\
  }\href {\doibase 10.1021/jp2122888} {\bibfield  {journal} {\bibinfo
  {journal} {J. Phys. Chem. C}\ }\textbf {\bibinfo {volume} {116}},\ \bibinfo
  {pages} {7546} (\bibinfo {year} {2012})}\BibitemShut {NoStop}%
\bibitem [{\citenamefont {Huang}\ \emph {et~al.}(2015)\citenamefont {Huang},
  \citenamefont {Xiao},\ and\ \citenamefont {Gao}}]{Huang2015}%
  \BibitemOpen
  \bibfield  {author} {\bibinfo {author} {\bibfnamefont {Y.}~\bibnamefont
  {Huang}}, \bibinfo {author} {\bibfnamefont {J.~J.}\ \bibnamefont {Xiao}}, \
  and\ \bibinfo {author} {\bibfnamefont {L.}~\bibnamefont {Gao}},\ }\href
  {\doibase 10.1364/OE.23.008818} {\bibfield  {journal} {\bibinfo  {journal}
  {Opt. Express}\ }\textbf {\bibinfo {volume} {23}},\ \bibinfo {pages} {8818}
  (\bibinfo {year} {2015})}\BibitemShut {NoStop}%
\bibitem [{\citenamefont {Baranov}\ \emph {et~al.}(2013)\citenamefont
  {Baranov}, \citenamefont {Andrianov}, \citenamefont {Vinogradov},\ and\
  \citenamefont {Lisyansky}}]{Baranov}%
  \BibitemOpen
  \bibfield  {author} {\bibinfo {author} {\bibfnamefont {D.~G.}\ \bibnamefont
  {Baranov}}, \bibinfo {author} {\bibfnamefont {E.~S.}\ \bibnamefont
  {Andrianov}}, \bibinfo {author} {\bibfnamefont {A.~P.}\ \bibnamefont
  {Vinogradov}}, \ and\ \bibinfo {author} {\bibfnamefont {A.~A.}\ \bibnamefont
  {Lisyansky}},\ }\href@noop {} {\bibfield  {journal} {\bibinfo  {journal}
  {Opt. Express}\ }\textbf {\bibinfo {volume} {21}},\ \bibinfo {pages}
  {10779–10791} (\bibinfo {year} {2013})}\BibitemShut {NoStop}%
\bibitem [{\citenamefont {Hao}\ and\ \citenamefont
  {Nordlander}(2007)}]{Hao2007}%
  \BibitemOpen
  \bibfield  {author} {\bibinfo {author} {\bibfnamefont {F.}~\bibnamefont
  {Hao}}\ and\ \bibinfo {author} {\bibfnamefont {P.}~\bibnamefont
  {Nordlander}},\ }\href {\doibase 10.1016/j.cplett.2007.08.027} {\bibfield
  {journal} {\bibinfo  {journal} {Chem. Phys. Lett.}\ }\textbf {\bibinfo
  {volume} {446}},\ \bibinfo {pages} {115} (\bibinfo {year}
  {2007})}\BibitemShut {NoStop}%
\bibitem [{\citenamefont {Stockman}(2011)}]{Stockman2010}%
  \BibitemOpen
  \bibfield  {author} {\bibinfo {author} {\bibfnamefont {M.~I.}\ \bibnamefont
  {Stockman}},\ }\href {\doibase 10.1103/PhysRevLett.106.156802} {\bibfield
  {journal} {\bibinfo  {journal} {Phys. Rev. Lett.}\ }\textbf {\bibinfo
  {volume} {106}},\ \bibinfo {pages} {156802} (\bibinfo {year} {2011})},\
  \Eprint {http://arxiv.org/abs/1011.3751} {arXiv:1011.3751} \BibitemShut
  {NoStop}%
\bibitem [{\citenamefont {Oldenburg}\ \emph {et~al.}(1998)\citenamefont
  {Oldenburg}, \citenamefont {Averitt}, \citenamefont {Westcott},\ and\
  \citenamefont {Halas}}]{Oldenburg1998}%
  \BibitemOpen
  \bibfield  {author} {\bibinfo {author} {\bibfnamefont {S.}~\bibnamefont
  {Oldenburg}}, \bibinfo {author} {\bibfnamefont {R.}~\bibnamefont {Averitt}},
  \bibinfo {author} {\bibfnamefont {S.}~\bibnamefont {Westcott}}, \ and\
  \bibinfo {author} {\bibfnamefont {N.}~\bibnamefont {Halas}},\ }\href
  {\doibase 10.1016/S0009-2614(98)00277-2} {\bibfield  {journal} {\bibinfo
  {journal} {Chem. Phys. Lett.}\ }\textbf {\bibinfo {volume} {288}},\ \bibinfo
  {pages} {243} (\bibinfo {year} {1998})}\BibitemShut {NoStop}%
\bibitem [{\citenamefont {Razink}\ and\ \citenamefont
  {Schlotter}(2007)}]{Razink2007}%
  \BibitemOpen
  \bibfield  {author} {\bibinfo {author} {\bibfnamefont {J.}~\bibnamefont
  {Razink}}\ and\ \bibinfo {author} {\bibfnamefont {N.}~\bibnamefont
  {Schlotter}},\ }\href {\doibase 10.1016/j.jnoncrysol.2007.06.067} {\bibfield
  {journal} {\bibinfo  {journal} {J. Non-Cryst. Solids}\ }\textbf {\bibinfo
  {volume} {353}},\ \bibinfo {pages} {2932} (\bibinfo {year}
  {2007})}\BibitemShut {NoStop}%
\bibitem [{\citenamefont {Jana}\ \emph {et~al.}(2001)\citenamefont {Jana},
  \citenamefont {Gearheart},\ and\ \citenamefont {Murphy}}]{Jana2001}%
  \BibitemOpen
  \bibfield  {author} {\bibinfo {author} {\bibfnamefont {N.~R.}\ \bibnamefont
  {Jana}}, \bibinfo {author} {\bibfnamefont {L.}~\bibnamefont {Gearheart}}, \
  and\ \bibinfo {author} {\bibfnamefont {C.~J.}\ \bibnamefont {Murphy}},\
  }\href {\doibase 10.1021/la0104323} {\bibfield  {journal} {\bibinfo
  {journal} {Langmuir}\ }\textbf {\bibinfo {volume} {17}},\ \bibinfo {pages}
  {6782} (\bibinfo {year} {2001})}\BibitemShut {NoStop}%
\bibitem [{\citenamefont {Malitson}(1965)}]{Malitson1965}%
  \BibitemOpen
  \bibfield  {author} {\bibinfo {author} {\bibfnamefont {I.~H.}\ \bibnamefont
  {Malitson}},\ }\href {\doibase 10.1364/JOSA.55.001205} {\bibfield  {journal}
  {\bibinfo  {journal} {J. Opt. Soc. Am.}\ }\textbf {\bibinfo {volume} {55}},\
  \bibinfo {pages} {1205} (\bibinfo {year} {1965})}\BibitemShut {NoStop}%
\bibitem [{\citenamefont {Johnson}\ and\ \citenamefont
  {Christy}(1972)}]{PhysRevB.6.4370}%
  \BibitemOpen
  \bibfield  {author} {\bibinfo {author} {\bibfnamefont {P.~B.}\ \bibnamefont
  {Johnson}}\ and\ \bibinfo {author} {\bibfnamefont {R.~W.}\ \bibnamefont
  {Christy}},\ }\href {\doibase 10.1103/PhysRevB.6.4370} {\bibfield  {journal}
  {\bibinfo  {journal} {Phys. Rev. B}\ }\textbf {\bibinfo {volume} {6}},\
  \bibinfo {pages} {4370} (\bibinfo {year} {1972})}\BibitemShut {NoStop}%
\bibitem [{\citenamefont {Suzuki}\ and\ \citenamefont
  {Lee}(2008)}]{Suzuki2008}%
  \BibitemOpen
  \bibfield  {author} {\bibinfo {author} {\bibfnamefont {H.}~\bibnamefont
  {Suzuki}}\ and\ \bibinfo {author} {\bibfnamefont {I.-y.~S.}\ \bibnamefont
  {Lee}},\ }\href@noop {} {\bibfield  {journal} {\bibinfo  {journal} {Int. J.
  Phys. Sci.}\ }\textbf {\bibinfo {volume} {3}},\ \bibinfo {pages} {038}
  (\bibinfo {year} {2008})}\BibitemShut {NoStop}%
\bibitem [{\citenamefont {Gao}\ \emph {et~al.}(2009)\citenamefont {Gao},
  \citenamefont {He}, \citenamefont {Deng}, ,\ and\ \citenamefont
  {Cao}}]{radhamine}%
  \BibitemOpen
  \bibfield  {author} {\bibinfo {author} {\bibfnamefont {X.}~\bibnamefont
  {Gao}}, \bibinfo {author} {\bibfnamefont {J.}~\bibnamefont {He}}, \bibinfo
  {author} {\bibfnamefont {L.}~\bibnamefont {Deng}}, , \ and\ \bibinfo {author}
  {\bibfnamefont {H.}~\bibnamefont {Cao}},\ }\href@noop {} {\bibfield
  {journal} {\bibinfo  {journal} {Opt. Mater.}\ }\textbf {\bibinfo {volume}
  {31}},\ \bibinfo {pages} {1715} (\bibinfo {year} {2009})}\BibitemShut
  {NoStop}%
\bibitem [{\citenamefont {Press}\ \emph {et~al.}(2007)\citenamefont {Press},
  \citenamefont {Teukolsky}, \citenamefont {Vetterling},\ and\ \citenamefont
  {Flannery}}]{Press:2007:NRE:1403886}%
  \BibitemOpen
  \bibfield  {author} {\bibinfo {author} {\bibfnamefont {W.~H.}\ \bibnamefont
  {Press}}, \bibinfo {author} {\bibfnamefont {S.~A.}\ \bibnamefont
  {Teukolsky}}, \bibinfo {author} {\bibfnamefont {W.~T.}\ \bibnamefont
  {Vetterling}}, \ and\ \bibinfo {author} {\bibfnamefont {B.~P.}\ \bibnamefont
  {Flannery}},\ }\href@noop {} {\emph {\bibinfo {title} {Numerical Recipes 3rd
  Edition: The Art of Scientific Computing}}},\ \bibinfo {edition} {3rd}\ ed.\
  (\bibinfo  {publisher} {Cambridge University Press},\ \bibinfo {year}
  {2007})\BibitemShut {NoStop}%
\bibitem [{\citenamefont {Rotter}(2009)}]{Rotter}%
  \BibitemOpen
  \bibfield  {author} {\bibinfo {author} {\bibfnamefont {I.}~\bibnamefont
  {Rotter}},\ }\href {http://stacks.iop.org/1751-8121/42/i=15/a=153001}
  {\bibfield  {journal} {\bibinfo  {journal} {J. Phys. A: Math. Theor.}\
  }\textbf {\bibinfo {volume} {42}},\ \bibinfo {pages} {153001} (\bibinfo
  {year} {2009})}\BibitemShut {NoStop}%
\end{thebibliography}%

\end{document}